\documentclass[prd,article,twocolumn,preprintnumbers,superscriptaddress,amsmath,amssymb,aps,longbibliography,nofootinbib]{revtex4-2}

\usepackage{dsfont}
\usepackage[utf8]{inputenc}
\usepackage{hyperref}
\usepackage{graphicx}   
\usepackage[table,dvipsnames]{xcolor}
\usepackage{colortbl}
\usepackage{ragged2e}
\usepackage{url}
\usepackage[normalem]{ulem}
\usepackage[english]{babel}
\usepackage{slashed,centernot}
\usepackage{bbold}
\usepackage[capitalise, english]{cleveref}
\usepackage[noindentafter]{titlesec} 
\usepackage{epigraph}
\usepackage{booktabs} 
\usepackage{upgreek}
\usepackage{physics}
\usepackage{pifont}
\usepackage{booktabs}
\usepackage{braket}
\usepackage{float}
\usepackage{soul}
\usepackage{slashed}
\usepackage{orcidlink}
\usepackage[compat=1.1.0]{tikz-feynman}

\definecolor{red}{rgb}{1,.0706,.1373}
\definecolor{blue}{rgb}{0,0.396,0.741}
\definecolor{realblue}{rgb}{0,0,1}
\definecolor{green}{rgb}{0.25,0.6,0.2}
\definecolor{rossoc}{cmyk}{0,1,1,0.2}

\definecolor{teal}{rgb}{0, 0,0.8}
\colorlet{mylinkcolor}{teal}
\colorlet{mycitecolor}{teal}
\colorlet{myurlcolor}{teal}
\hypersetup{
  linkcolor  = mylinkcolor!,
  citecolor  = mycitecolor!,
  urlcolor   = myurlcolor!,
  colorlinks = true
}
\newcommand{\hc}{\; + \; \mathrm{H.c.} \;}
\def\L{\mathcal{L}}

\begin{document}

\title{
The Majoron Cosmological Window:\\ Dark Matter and Thermal Leptogenesis
}

\preprint{IPPP/26/39}

\author{Arturo de Giorgi~\orcidlink{0000-0002-9260-5466}
}
\email{arturo.de-giorgi@durham.ac.uk}
\affiliation{Institute for Particle Physics Phenomenology, Department of Physics, Durham University, Durham DH1 3LE, U.K.}

\author{Daniel Naredo-Tuero~\orcidlink{0000-0002-5161-5895}}
\email{daniel.naredo@kit.edu}
\affiliation{Instituto de F\'{\i}sica Te\'orica UAM/CSIC, Universidad Aut\'onoma de Madrid, Cantoblanco, 28049 Madrid, Spain}
\affiliation{Institute for Astroparticle Physics (IAP), Karlsruhe Institute of Technology (KIT), Hermann-von-Helmholtz-Platz 1, 76344 Eggenstein-Leopoldshafen, Germany}

\author{Xavier Ponce D\'iaz~\orcidlink{0000-0002-1305-1187}}
\email{xavier.poncediaz@unibas.ch}
\affiliation{Department of Physics, University of Basel, Klingelbergstrasse 82,  CH-4056 Basel, 
Switzerland}

%=============================================================================

\begin{abstract}
The majoron is the Nambu-Goldstone boson associated with the spontaneous breaking of a global $B-L$ symmetry. Remarkably, the minimal majoron framework can simultaneously address three key empirical indications of physics beyond the Standard Model: neutrino masses, the matter-antimatter asymmetry, and dark matter. In this work, we identify the cosmologically viable region in which majoron dark matter and high-scale thermal leptogenesis can be realised simultaneously. We show that successful leptogenesis plays a central role in making this scenario predictive: by constraining the right-handed-neutrino mass scale, it determines the irreducible freeze-in contribution to the majoron abundance and fixes the size of the couplings relevant for visible dark matter decays. Combining the irreducible dark matter production mechanisms with warm dark matter limits and indirect searches for decaying dark matter, we map the resulting majoron cosmological window and show that future X- and gamma-ray telescopes can probe part of the surviving parameter space. 
\end{abstract}
\maketitle

%-----------------------------------------------------------------------------
\tableofcontents
%-----------------------------------------------------------------------------

\section{Introduction} 
\label{sec:intro}
%-----------------------------------------------------------------------------
Among the leading empirical indications of physics beyond the Standard Model~(SM), dark matter~(DM), neutrino masses, and the observed matter–antimatter asymmetry require the extension of the SM particle content. While the latter two can be minimally explained within the framework of the type-I seesaw~\cite{Minkowski:1977sc, Yanagida:1980xy, Gell-Mann:1979vob, Mohapatra:1979ia, Schechter:1980gr, CentellesChulia:2024uzv} and (thermal) leptogenesis mechanisms~\cite{Fukugita:1986hr,Covi:1996wh,Covi:1996fm,Pilaftsis:1997jf,Buchmuller:1997yu}, the existence of DM remains unaccounted for in this minimal setup.

In the landscape of proposed SM extensions~\cite{Albertus:2026fbe,Arza:2026rsl}, the \textit{majoron}~\cite{Chikashige:1980ui, Chikashige:1980qk, Schechter:1981cv, Gelmini:1980re} is arguably one of the simplest frameworks capable of accommodating DM. The majoron is the pseudo-Nambu--Goldstone boson (pNGB) associated with the spontaneous breaking of a global $B-L$ symmetry. Its interactions are suppressed by the seesaw scale, making it naturally weakly coupled to matter and stable on cosmological timescales. These properties render the majoron a natural DM candidate.

The pNGB nature of the majoron allows it to be naturally light. Its mass, however, is not predicted in the minimal theory. Although several mechanisms can generate it, cf.~\cite{Rothstein:1992rh, Nomura:2000yk,Frigerio:2011in,deGiorgi:2023tvn,Greljo:2025suh,Berbig:2025nrt},  its origin is not tied to the solution of a theoretical puzzle, such as the way the axion mass is linked to the strong-CP problem~\cite{Weinberg:1977ma,Wilczek:1977pj,Peccei:1977hh,Peccei:1977ur}. Nevertheless, global symmetries are expected to be broken at least by gravitational effects, making a massive majoron an appealing possibility. From a bottom-up perspective, the majoron mass can therefore be treated as a free parameter, to be fixed by the requirement of reproducing the observed DM relic abundance.\footnote{Note that under the assumption of a gravity-induced breaking, ``naively'', operators arising at small dimension may lead to several problems, see Refs.~\cite{Akhmedov:1992hi,Cline:1993ht}. Hence, some sort of protection needs to be imposed~\cite{Rothstein:1992rh,Greljo:2025suh}.}

The ultra-weak couplings of the majoron, rooted in the high seesaw scale, are both a virtue and a limitation. On the one hand, the resulting suppression of its interactions with light fermion families and photons~\cite{Heeck:2019guh,Herrero-Brocal:2023czw} allows it to evade most astrophysical constraints, rendering it a naturally \textit{astrophobic}\footnote{For examples of models of astrophobic axions, see Refs.~\cite{DiLuzio:2017ogq,Bjorkeroth:2019jtx,DiLuzio:2022tyc}.} Axion-Like Particle (ALP). On the other hand, the same suppression makes its experimental detection challenging. Nevertheless, majoron DM remains a prime target for neutrino, X-ray, and gamma-ray telescopes~\cite{Garcia-Cely:2017oco,Akita:2023qiz}. Moreover, extensions of the minimal majoron framework can make it accessible to a broader range of experimental searches~\cite{Arias-Aragon:2020qip,Arias-Aragon:2022ats,Liang:2024vnd,Lu:2025kbe}.

However, the presence of the majoron can interplay non-trivially with different leptogenesis mechanisms, which, for simplicity, we can divide into three classes: scalar-induced, low-scale, and high-scale leptogenesis:
\begin{itemize}
    \item Coherent majoron motion can realize spontaneous leptogenesis~\cite{Cohen:1987vi,Cohen:1988kt}, acting as an effective chemical potential for lepton number~\cite{Ibe:2015nfa,Chun:2023eqc,Wada:2024cbe,Chun:2025abp,Takahashi:2026ngu}; related Affleck--Dine-type constructions~\cite{Affleck:1984fy} use the lepton-number-carrying scalar condensate to store the asymmetry before transferring it to the thermal bath~\cite{Mohapatra:2021ozu}. These scenarios do not arise generically in the minimal majoron framework, but instead rely on additional assumptions such as large majoron masses, potentially in tension with DM stability\footnote{A similar discussion applies to triplet majoron models, e.g. see Ref.~\cite{Berbig:2025hlc}.}, sizeable primordial angular motion, or extra model-building ingredients.
    \item  In low-scale majoron models, leptogenesis can be achieved via the Akhmedov-Rubakov-Smirnov~\cite{Akhmedov:1998qx,Asaka:2005pn} mechanism~\cite{Caputo:2018zky,Escudero:2021rfi}, requiring masses of the right-handed neutrinos~(RHNs) $M_N \sim 1-100\,$GeV.
    \item In the scenario of standard high-scale leptogenesis, this discussion is only found in analyses that introduce extra gauge bosons~\cite{Greljo:2025suh}, new $B-L$-violating operators~\cite{Brune:2022vzd,Manna:2022gwn,King:2024idj}, non-minimal extensions~\cite{Brune:2025zwg}, or new scalar interactions~\cite{Vilja:1993uw,Gu:2009hn,Gu:2010ys,AristizabalSierra:2014uzi}.
    These scenarios are qualitatively different from the one considered here. 
    A thermalised majoron decouples as a hot relativistic relic and is therefore constrained to be either very light or a subdominant DM component~\cite{Reig:2019sok}.
    Moreover, in the minimal majoron setup, thermalisation is not generic: direct couplings to the SM are seesaw/loop-suppressed, while efficient equilibration requires either sizeable RHN couplings or a Higgs--singlet portal involving the radial mode. 
\end{itemize}

\newpage
Motivated by the region of parameter space accessible to current and future searches for majoron DM~\cite{Garcia-Cely:2017oco,Akita:2022lit}, we study the irreducible DM production mechanisms of the minimal majoron model and their compatibility with high-scale thermal leptogenesis and experimental searches. To the best of our knowledge, a systematic analysis of this combined cosmologically viable window has not appeared in the literature.\footnote{\textbf{Note added:} during the very last stage of writing of this manuscript, Ref.~\cite{Akita:2026gzk} appeared, addressing the same scenario with a complementary analysis. Our work provides an independent assessment of the viable parameter space, with a detailed scan of the seesaw parameters and a unified treatment of relic-density, leptogenesis, warm-DM, and decay constraints.} Although standard leptogenesis may proceed as in the type-I seesaw, its simultaneous realisation with majoron DM has two important consequences that have not been fully incorporated in previous analyses. First, high-scale thermal leptogenesis requires the right-handed neutrinos to be thermally populated, or at least sufficiently close to equilibrium~\cite{Giudice:2003jh}; through the renormalisable majoron--RHN coupling, this inevitably generates a freeze-in population of majorons~\cite{Heeck:2017xbu,Boulebnane:2017fxw,Greljo:2025suh}, which can be relevant for keV-scale DM. Second, the requirement of successful leptogenesis provides an independent handle on the heavy-neutrino mass scale, thereby lifting otherwise flat directions in the majoron parameter space and making decay bounds into charged fermions and photons predictive. 

In this work, we analyze the interplay of these three empirical indications of physics beyond the Standard Model within the minimal majoron framework. This allows us to define a \textit{cosmologically viable} window for majoron models, which can be targeted by future DM searches. We focus on cosmological histories in which the early Universe is sufficiently hot for the right-handed neutrinos to participate in the thermal plasma. This provides the common origin of the two central ingredients of our analysis: their out-of-equilibrium dynamics generate the matter--antimatter asymmetry through thermal leptogenesis, while their unavoidable interactions with the majoron produce an irreducible contribution to the DM abundance.

%%%%%%%%%%%%%%%%%%%%%%%%%%%%%%%%%%%%%%%%%%%%
%%%%%%%%%%%%%%%%%%%%%%%%%%%%%%%%%%%%%%%%%%%%
\section{The Model}
The majoron's minimal embedding consists of a complex scalar singlet whose vacuum expectation value generates the mass scale of light and heavy RHNs through the seesaw mechanism. The Lagrangian of the theory is given by 
\begin{equation}
    \label{eq:maj_lag}
    -\L_\phi = \bar{L}_L\, Y_D \tilde{H} N_R +  \phi \bar{N}_R^c Y_N N_R + V(\phi)  \hc,
\end{equation}
where $L_L$ is the leptonic weak-doublet, $H$~($\tilde{H}$) is the Higgs (dual), $N_R$ is a right-handed~(RH) neutrino, and $\phi$ is the complex scalar singlet which gives rise to the majoron. In the above Lagrangian, we omitted flavour indices; we will assume the existence of three RH neutrinos, so that $Y_{D,N}$ are $3\times3$ matrices. Furthermore, $Y_N$ can be chosen to be real and diagonal without loss of generality. 

The majoron, $a$, can be identified as the angular mode of the complex singlet field~\footnote{The mass of the radial mode of the singlet is expected to be of $\mathcal{O}(f_a)$ and can be made heavier by choosing the parameters of the potential. It is typically neglected as its phenomenology and impact strongly depend on the choice of scalar potential. We will therefore not consider it in these studies, as commonly done in the literature.} 
\begin{align}
    &\phi = \frac{(f_a+\rho)}{\sqrt{2}} e^{i a/f_a}\, , &&\textrm{ with }  &&\left<|\phi|^2\right>=\frac{f_a^2}{2} \, .
\end{align}  We will be agnostic about the singlet's potential $V(\phi)$, assuming only the presence of a small soft-breaking of the $\mathrm{U}(1)$ which will give rise to the majoron mass.
All in all, the effective majoron Lagrangian can be written as
\begin{equation}
    \L_\mathrm{majoron}=\frac{1}{2}(\partial_\mu a)^2-\frac{1}{2}m_a^2 a^2 +\mathcal{L}_{a\text{-SM}}\,.
\end{equation}
where $\mathcal{L}_\text{a-SM}$ encodes the interactions between the majoron and the SM. While throughout our study we will consider $m_a$ and $f_a$ to be independent parameters, once a source of explicit $\mathrm{U}(1)$-breaking is specified, they become related. For illustrative purposes, we will show $m_a-f_a$ relationships stemming from gravity-induced $\mathrm{U}(1)$-breakings of the form
\begin{equation}
    \mathcal{L}_{\mathrm{U}(1)-\rm break}=\left(4\pi\right)^2\frac{\eta}{M_{\rm Pl}^{n-4}}\phi^n+{\rm h.c.}\, ,
    \label{eq:explicit_breaking_terms}
\end{equation}
where $\eta$ is typically an $\mathcal{O}(1)$ number. This operator leads to a majoron mass given by:
\begin{equation}
\label{eq:mass_planck}
    m_a^2=16\pi^2n^2|\eta|M_{\rm Pl}^2\left(\frac{f_a}{\sqrt{2} M_{\rm Pl}}\right)^{n-2} \, .
\end{equation}
Lines corresponding to $m_a-f_a$ relations stemming from different dimensions ($n$) of such explicit breaking are shown in Figs.~\ref{fig:parameter_space_post} and~\ref{fig:parameter_space_pre}.

The SSB generates the heavy sterile neutrinos' masses $M_{N,i}\simeq Y_{N,i}f_a/\sqrt{2}$. The active light neutrino masses $m_i$ are then generated via Type-I seesaw
\begin{equation}
    \text{diag}(m_1,m_2,m_3)\simeq- U_\text{PMNS}^T\left(M_D M_N^{-1}M_D^T \right)U_\text{PMNS}\,,
\end{equation}
with $M_D\equiv Y_D\, v /\sqrt{2}$, and $v=246\,$GeV, the Electroweak (EW) vacuum expectation value (vev). While the two light neutrino mass splittings ($\Delta m_{21}^2$ and $\Delta m^2_{31}$) are precisely measured in neutrino oscillation experiments~\cite{Esteban:2024eli}, both the ordering of neutrino mass eigenstates and the lightest neutrino mass are yet unknown. For the sake of simplicity, throughout our analysis we will assume the so-called \textit{normal ordering} of neutrino masses ($m_1<m_2<m_3$) and a conservative upper limit on the lightest neutrino mass of $m_1<0.2\,\rm eV$.

The majoron couples at tree level only to neutrinos, while its interactions with charged SM fermions and gauge bosons are generated radiatively. In particular, the leading couplings to charged leptons arise at one loop, whereas those to photons and gluons first appear at two loops~\cite{Heeck:2019guh,Herrero-Brocal:2023czw}. We collect the relevant interactions in App.~\ref{app:couplings}. As expected for a pNGB, the fermionic couplings are proportional to the corresponding fermion masses. Nevertheless, over much of the parameter space, the neutrino coupling remains the dominant one: in particular, it exceeds the electron coupling for $f_a \lesssim 10^{10}$~GeV, see \cref{eq:elec_vs_neutrino}. The photon coupling is further suppressed by the electromagnetic coupling, by the additional loop factor, and, for light majorons, by powers of the majoron mass. Consequently, it vanishes in the massless limit and becomes phenomenologically relevant only for $m_a \sim \mathcal{O}(\mathrm{MeV})$ or larger.

%%%%%%%%%%%%%%%%%%%%%%%%%%%%%%%%
\section{Scrutinising the Parameter Space}
\label{sec:ParameterSpace}
%&&&&&&&&&&&&&&&&&&&&&&&&&&&&&&&&&&&&&&&&&&&&&&&&&

 %-----------------------------------------------------------------------------
\subsection{Experimental Constraints}
\label{sec:constraints}

Within the framework of the type-I seesaw, the relation between the Dirac Yukawas $Y_D$ and the seesaw scale $M_N = Y_N f_a/\sqrt{2}$ is tightly linked to the mass of the neutrinos. Furthermore, for light majorons, the photon coupling gets extremely suppressed. Therefore, generic constraints on sub-MeV ALPs from stellar evolution~\cite{Capozzi:2020cbu,Dolan:2022kul} or supernovae dynamics~\cite{Farzan:2002wx,Brune:2018sab,Caputo:2022mah,Diamond:2023scc,Fiorillo:2022cdq} do not provide relevant bounds for our scenario. More importantly, cosmological constraints in this regime, such as those arising from modifications of the primordial abundances produced during Big Bang Nucleosynthesis or from distortions of the CMB anisotropy spectra~\cite{Cadamuro:2011fd,Millea:2015qra,Depta:2020wmr,EscuderoAbenza:2025tsi}, are naturally evaded owing to the long lifetime of the high-scale majoron. For relevant cosmological bounds and effects of a low-scale $B-L$ majoron see Refs.~\cite{Escudero:2019gvw,Escudero:2021rfi}.

The dominant bounds on Majoron DM arise from its SM decay channels, the most relevant of which are neutrinos, charged fermions, and photons.
The majoron decay width into neutrinos is controlled by the ratio $m_\nu/f_a$ and is given by
\begin{equation}
    \Gamma_{a\to\nu\nu}=\frac{m_a\sum_{i=1}^3 m_i^2}{16\pi f_a^2}\,,
    \label{eq:neutrino_decay_width}
\end{equation}
where $\left\{m_i\right\}_{i=1}^3$ are the light neutrino masses. Neutrino observatories looking for neutrino lines with energy $E_\nu=m_a/2$ are able to directly constrain the $m_a-f_a$ parameter space for $m_a\gtrsim\rm few\,MeV$~\cite{Super-Kamiokande:2011lwo, Super-Kamiokande:2013ufi,Albert:2016emp,Garcia-Cely:2017oco,Borexino:2019wln,KamLAND:2021gvi,Super-Kamiokande:2021jaq,IceCube:2021kuw,Akita:2022lit,Arguelles:2022nbl,IceCube:2023ies,Akita:2023qiz}.
\begin{figure*}
    \centering
    \includegraphics[width=0.9\linewidth]{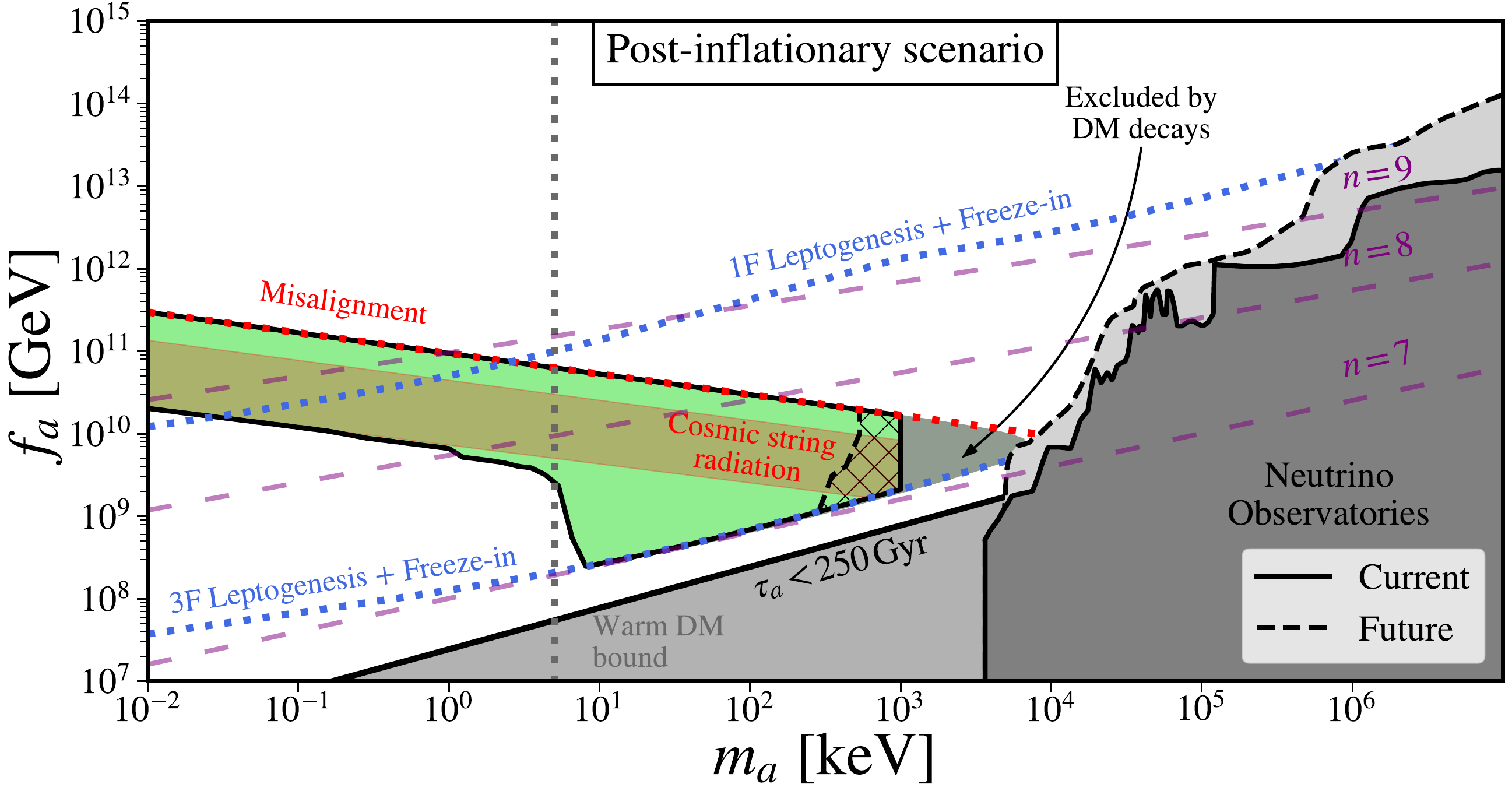}
    \caption{
    Parameter space for majoron DM in the post-inflationary scenario, compatible with successful thermal leptogenesis. {For this scenario to be viable, symmetry restoration cannot be thermal, see text for details}. The green region denotes the cosmologically viable window, where the observed DM abundance can be obtained from the irreducible misalignment and freeze-in contributions while satisfying DM decay and warm-DM bounds. {Note that wDM are applied consistently, taking from Ref.~\cite{Tan:2024cek} bounds on the fraction of wDM; however, in the plot we mark the commonly used bound of $5\, \textrm{keV}$ for reference}. Current and projected neutrino-telescope constraints are shown in dark-gray and light gray, respectively. {The other gray region is excluded by searches for decaying DM into charged particles and photons, while the dashed contour with hatched pattern indicates the projected sensitivity of future satellite missions~\cite{Thorpe-Morgan:2020rwc,Shutt:2025xvc}}. The red band shows the expected contribution from DM production via cosmic-string radiation. Purple dashed lines show the prediction of $m_a - f_a$ relations, if the mass is given by Planck-induced operators, \cref{eq:mass_planck}.
    }
    \label{fig:parameter_space_post}
\end{figure*}

For the decay to SM fermions and photons, the size of the couplings (see App.~\ref{app:couplings}) is instead controlled by the entries of $M_DM_D^\dagger/(vf_a)\,,$
which, parametrically, is $\mathcal{O}\left(\tfrac{m_\nu}{v}\tfrac{M_N}{f_a}\right)$. 
The dependence of the decays on the heavy neutrinos' Yukawas means that bounds on these decay channels can only be translated to $m_a-f_a$ space if $M_N$ can be independently constrained. 
On the other hand, the requirement of successful thermal leptogenesis does %as it 
provide a lower bound on the RHN mass scale, thus allowing us to translate the constraints on $M_D M_D^\dagger/(v f_a)$ stemming from majoron DM decay searches into lower bounds on $f_a$.

{On the astrophysical side, as we will later see, the strongest constraints on viable majorons come from decays to photons and electrons.} Photons produced from decaying DM are being targeted by $X$- and gamma-ray observatories~\cite{Perez:2016tcq,Ng:2019gch,Thorpe-Morgan:2020rwc,Foster:2021ngm,Dekker:2021bos,Roach:2022lgo,Calore:2022pks,Fong:2024qeq,Shutt:2025xvc}, providing the best constraints in the $\textrm{keV}-100\,\textrm{MeV}$ for ALP-DM, cf.~\cite{Queiroz:2014yna,AxionLimits}.

On the cosmological side, majoron DM decays inject energy into the early Universe and therefore alter both the ionisation and temperature histories. Measurements of the CMB anisotropies power spectra are highly sensitive to changes in the free electron fraction~\cite{Adams:1998nr,Padmanabhan:2005es,Slatyer:2009yq,Poulin:2016anj,Slatyer:2016qyl}, while Lyman-$\alpha$ observations~\cite{Baur:2015jsy, Yeche:2017upn, Baur:2017stq, Irsic:2017ixq} constrain the temperature of the intergalactic medium around the reionisation epoch. In the near future, measurements of the 21-cm signal will also be powerful probes of decaying DM~\cite{Sun:2023acy}. Additionally, cosmology is also sensitive to invisible decays, as is the case for sub-MeV majorons, which almost exclusively decay to a pair of neutrinos. By a combination of CMB+LSS data, the DM lifetime can be constrained to be $\tau>250\,\rm Gyr$~\cite{Simon:2022ftd}, as decaying DM not only affects the expansion history of the Universe, but also the growth of structure.

Finally, warm DM (wDM) constraints apply. If DM is too relativistic, it suppresses structure formation at small scales.
If the totality of DM is warm, this effectively translates into a lower bound on the wDM mass, $m_{\rm wDM}\gtrsim 5\,{\rm keV}$~\cite{Baur:2015jsy,Yeche:2017upn,Baur:2017stq,Irsic:2017ixq}. However, since majoron DM is generically a mixture of cold and warm components (see Sec.~\ref{sec:darkmatter}), we will employ mixed DM bounds~\cite{Kamada:2016vsc,Schneider:2018xba,Parimbelli:2021mtp,Tan:2024cek,An:2025gju}, which allow to constrain the relative abundance of wDM as a function of $m_{\rm wDM}$. 

In our analysis, we will consider the compilation of cosmological and astrophysical majoron bounds given in Ref.~\cite{Akita:2023qiz}, as well as additional bounds on the decay to photons below the MeV~\cite{AxionLimits,Thorpe-Morgan:2020rwc,Shutt:2025xvc} and the recent bounds on the decay to electrons provided in Ref.~\cite{Nguyen:2025tkl}.

%-----------------------------------------------------------------------------
\subsection{Majoron Dark Matter} 
\label{sec:darkmatter}

The feebly interacting nature of the majoron easily renders it cosmologically stable, making it a natural DM candidate. Generically, there exist three irreducible main mechanisms that contribute to generating its relic abundance\footnote{For early discussions on thermalised DM majoron see~\cite{Lattanzi:2007ux,Lattanzi:2008ds}.}: via misalignment, by the freeze-in mechanism, and by decay of topological defects related to the $\mathrm{U}(1)$ breaking.

The misalignment mechanism~\cite{Abbott:1982af,Dine:1982ah,Preskill:1982cy}, provides a non-thermal mechanism for cold DM production. The scalar field evolution in the background expanding universe, leads to the equation of motion:
\begin{equation}
    \ddot{\theta}+3H\dot{\theta}+m_a^2(t)\theta\simeq0\,.
\end{equation}
with $\theta\equiv a/f_a$. At high temperatures, the field is fixed by Hubble friction at its initial value $\theta=\theta_i$. However, as the Universe expands and cools down, the Hubble rate becomes comparable to the mass, $H(T_{\rm osc})=m_a$, and the field starts oscillating around the minimum of its potential. These oscillations redshift as matter and therefore contribute to the DM abundance.

The precise determination of DM abundance from misalignment depends on details of the model, such as the temperature dependence of the majoron mass, determined by the source of explicit $\mathrm{U}(1)$-breaking. %Nevertheless, if one assumes a 
For a temperature-independent majoron mass, $m_a(T)\simeq m_a$, its relic abundance is given by~\cite{Blinov:2019rhb}:
\begin{align}
    \label{eq:omega_misalignment}\Omega_{\rm DM}^{\rm mis}h^2\simeq&\,0.12~\left(\dfrac{m_a}{100\,{\rm k eV}}\right)^{1/2}\times\\
    &\nonumber \times\left(\dfrac{\theta_if_a}{3.38\times10^{10}\,{\rm GeV}}\right)^2\left(\dfrac{90}{g_*(T_{\rm osc})}\right)^{1/4}\, ,
\end{align}
where $g_\star(T)$ is the effective number of relativistic degrees
of freedom. We note that a $T$-independent majoron mass is justified in scenarios in which the source of explicit $\mathrm{U}(1)_{B-L}$ breaking, providing mass for the majoron, is associated to a scale $\Lambda\gg f_a$, above the scale of spontaneous $B-L$ breaking, which is the case with the terms we consider in Eq.~\eqref{eq:explicit_breaking_terms}.
Since misalignment is a non-thermal process, its contribution to the DM abundance is cold, allowing extremely light fields to constitute the DM.

The abundance of non-thermally produced DM crucially depends on the cosmological history of $\mathrm{U}(1)$ spontaneous breaking, namely, on whether the symmetry is broken after or before the end of inflation. These two scenarios are typically dubbed \textit{post-inflationary} and \textit{pre-inflationary}, respectively. In the former scenario, since the symmetry is restored after inflation, each Hubble patch was populated with a different, uncorrelated value of the initial misalignment angle, leading to an average value of $\theta_i=\sqrt{\tfrac{1}{2\pi}\int_{-\pi}^\pi d\theta \theta^2} =\tfrac{\pi}{\sqrt{3}}$. On the contrary, in the latter scenario, the symmetry is already broken before or during inflation, in which case a homogeneous patch with a random $\theta_i$ is inflated, meaning that $\theta_i$ is effectively an extra free parameter taking value in the interval $\left\lvert{\theta_i}\right\rvert\in [0,\pi]$. 

\bigskip
A second contribution to the relic majoron DM abundance comes from thermal production.
%can also occur thermally. 
Indeed, reproducing the BAU via thermal leptogenesis requires the existence of a thermal bath of at least one of the right-handed neutrinos and, since the majoron has sizable Yukawa couplings to RHN, a non-zero irreducible thermal abundance of majorons is guaranteed to be produced in this scenario. The freeze-in abundance can be obtained by integrating the Boltzmann equation for the majoron yield $Y_a$. Including production from $N_iN_i\to aa$ and from $HL\to N_i a$, $HN_i\to L a$ and  $L N_i\to H a$ (see App.~\ref{app:majoron_therm}), and following Ref.~\cite{Blennow:2013jba}, we find the parametric scaling
%\footnote{We assume $T_{\rm rh}\gtrsim M_{N_i}$ for the RH neutrino species relevant to freeze-in and leptogenesis, so that $N_i$ is thermally populated and the freeze-in production, dominated at $T\sim M_{N_i}$, takes place.}\DN{Given the red text, now this footnote might be redundant.}
\begin{equation}
\label{eq:omega_fi_estimation}
\begin{aligned}
    \Omega_{\rm DM}^{\rm fi}h^2 &\simeq
    0.12 \left(\frac{m_a}{100\,{\rm keV}}\right)
    \left(\frac{3.4\times 10^{10}\,{\rm GeV}}{f_a}\right)^2
    \times \\
    &\quad \times \sum_{i=1}^3
    \left[
    \left(\frac{3.4\times10^{10}\,{\rm GeV}}{f_a}\right)^2
    \left(\frac{M_{N_i}}{1.6\times 10^{8}\,{\rm GeV}}\right)^3
    \right. \\
    &\quad\left.
    +
    \left(\frac{\tilde{m}_i}{0.1\,{\rm eV}}\right)
    \left(\frac{M_{N_i}}{6.2\times 10^{8}\,{\rm GeV}}\right)^2
    \right] .
\end{aligned}
\end{equation}
where $\tilde m_i \equiv (M_D^\dagger M_D)_{ii}/M_{N_i} \sim \mathcal{O}(m_\nu)$. {We have assumed that the reheating temperature is large enough to generate a thermal bath of all three RHN, i.e. $T_{\rm rh}>M_{N_i}$. 
%While this condition is generally satisfied when the $U(1)$-breaking is post-inflationary, since it typically requires $T_{\rm rh}>f_a$, it may not be the case in pre-inflationary scenarios. 
Since we will later choose a mildly hierarchical RHN mass spectrum, the condition $T_{\rm rh}>M_{N_i}$ is approximately equivalent to the condition $T_{\rm rh}>M_{N_1}$, necessary for a successful thermal leptogenesis, which is one of the main assumptions of our analysis.}

It should be remarked that sizable majoron DM production from the thermal bath could also occur via the resonant $s$-channel $NN\to\rho\to a a$. 
%Since $m_\rho\simeq \mathcal{O}(f_a)$, these channels are relevant only if $T_{\rm rh}>f_a$. 
However, we argue that even in this case, this contribution is expected to be small compared with the $t$-channel since, in general, the freeze-in regimes require $M_N/f_a\ll1$ so as to preclude the thermalisation of the majoron, see Appendix~\ref{app:majoron_therm}. Therefore, in order for the $s$-channel to be relevant, it is necessary to have a thermal bath of RHN at $T\sim m_\rho\simeq f_a\gg M_N$. However, this is generally not possible in our minimal majoron setup, since, in the absence of additional interactions, RHN are populated from inverse decays of the leptonic and Higgs doublets $LH\to N$ which acquire $\mathcal{O}(T)$ thermal masses, effectively closing off the production of RHN for $T\gg M_N$. 

\begin{figure*}
    \centering
    \includegraphics[width=0.9\linewidth]{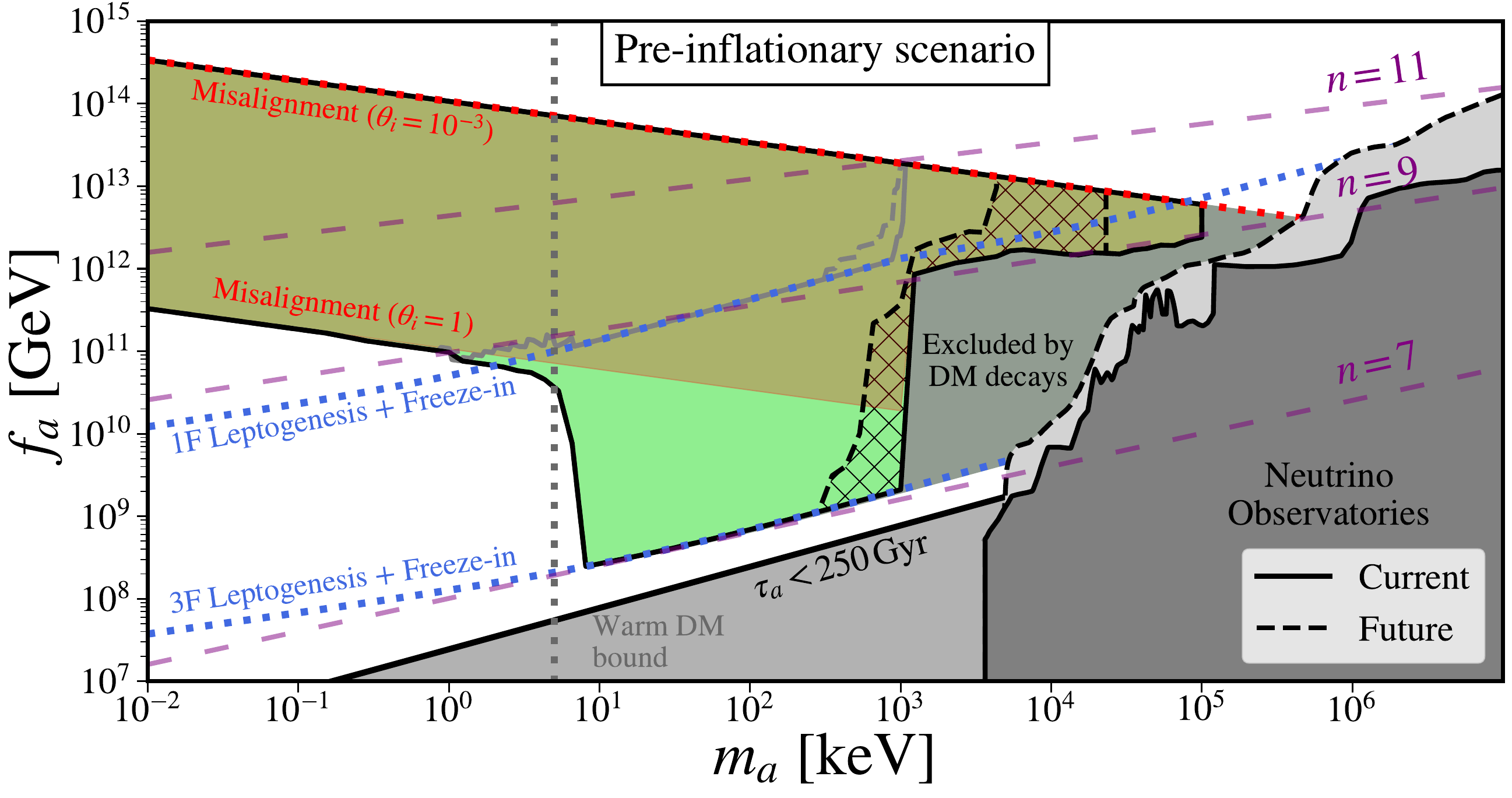}
    \caption{Parameter space for pre-inflationary Majoron DM compatible with successful thermal leptogenesis. 
    The green region shows the cosmologically viable window, and the red area the DM relic abundance dependence on the free initial misalignment angle $\theta_i$. 
    The dark-gray and light-gray regions show current and projected neutrino-observatory constraints, respectively. 
    {The other gray region indicates existing bounds from majoron decays into charged particles and photons, while the dashed lines with hatched region show the corresponding future sensitivities}. Purple lines show the $m_a-f_a$ prediction for different Planck-induced operators, \cref{eq:mass_planck}. {Bounds on the fraction of wDM have been used using Ref.~\cite{Tan:2024cek}, however, we mark the usual bound of wDM $5\,\textrm{keV}$ as a reference.}}
    \label{fig:parameter_space_pre}
\end{figure*}

 \bigskip
Finally, the spontaneous breaking of a $\mathrm{U}(1)$ symmetry inevitably leads to the formation of topological defects~\cite{Kibble_1976}, namely cosmic strings, which, if unstable as in our case, lose energy by radiating ALPs~\cite{DAVIS1986225}. This extra non-thermal contribution is absent in the pre-inflationary scenario, since the abundance of topological defects formed during inflation will be exponentially diluted. The contribution of string radiation to the DM abundance is an active field of research, mostly relying on computationally expensive cosmological simulations that track the dynamics of string formation, interaction, and decay. 
It is well known that the contribution to the DM abundance from string radiation exhibits a parametric dependence similar to that of misalignment, differing only in a numerical prefactor. Furthermore, both, analytical estimates~\cite{Reig:2019sok,Chathirathas:2025aan} and numerical simulations~\cite{Gorghetto:2018myk,Gorghetto:2020qws,Buschmann:2021sdq,Kim:2024dtq,Saikawa:2024bta,Benabou:2024msj,Kim:2024wku} point towards string radiation providing the larger contribution.

{Additionally, since in the pre-inflationary scenario the majoron is already present during inflation, its quantum fluctuations can then lead to isocurvature perturbations with a nearly scale-invariant power spectrum~\cite{Lyth:1991ub,Beltran:2006sq,Hertzberg:2008wr,Arvanitaki:2019rax} and an amplitude given by:
\begin{equation}
    \Delta^2_a(k)=\frac{H_I^2}{\pi^2\theta_i^2f_a^2}\equiv A_sf_{\rm iso}^2\,,
\end{equation}
where $H_I$ is the Hubble rate at the end of inflation, while $f_{\rm iso}$ is the amplitude of isocurvature perturbations normalised to that of adiabatic perturbations $A_s$. CMB observations constrain $f_{\rm iso}<0.2$, which translates into an upper bound on the energy scale of inflation $E_I$:
\begin{equation}
    E_I<3.5\times 10^{14}\,{\rm GeV}\sqrt{\frac{\theta_if_a}{10^{15}\,{\rm GeV}}}\,,
    \label{eq:isocurvature_bound}
\end{equation}
with $H_I$ and $E_I$ related by the Friedman equation $H_I=\sqrt{\tfrac{8\pi}{3}}\tfrac{E_I^2}{M_{\rm Pl}}$. Since $E_I$ provides an upper bound on the reheating temperature $T_{\rm rh}\leq E_I$, this bound must be taken into account, as our assumptions about the success of leptogenesis and freeze-in production require $T_{\rm rh}>M_{N_i}$. These bounds have been recently discussed in the context of majorons in Ref.~\cite{Batell:2026avi}.
}

 There is one last remark that needs to be discussed regarding the restoration of the $\mathrm{U}(1)$ symmetry in the post-inflationary scenario. In our setup, the symmetry restoration cannot be thermal, since the scalar $\phi$ only interacts with RHN, and we have argued that it is not possible to generate a thermal RHN bath at $T\gg M_N$. Furthermore, additional $\phi$-interactions that may lead to a thermal restoration can potentially lead to its thermalisation, which would prove catastrophic for DM production, as subsequent decays of the radial mode $\rho\to aa$ can easily equilibrate the majoron and lead to DM overproduction. Consequently, we consider that, in the post-inflationary scenario, the symmetry is non-thermally restored, as it occurs in inflationary models in which the scalar is coupled to the inflaton~\cite{Ballesteros:2016euj,Bao:2022hsg,Sopov:2025hhr}.

We also neglect a renormalizable Higgs portal interaction, $\lambda_{H\phi}|H|^2|\phi|^2$. For $f_a\gg v$, a generic value of $\lambda_{H\phi}$ would induce a threshold correction $\delta m_H^2\sim \lambda_{H\phi} f_a^2$, thereby reintroducing a tuning in the electroweak scale. Radiative stability therefore selects the technically natural ultra-weak regime $\lambda_{H\phi}\lesssim v^2/f_a^2$, since the limit $\lambda_{H\phi}\to0$ corresponds to a decoupling of the Higgs and singlet sectors, due to an extended Poincar\'e symmetry~\cite{Volkas:1988cm,Foot:2013hna}. In this regime, the portal is also cosmologically irrelevant: the interaction rate for Higgs-mediated processes involving the radial mode remains well below the Hubble rate at the relevant temperatures, and hence the radial mode $\rho$ is not thermalised through the Higgs portal.

Note that for a majoron scale $f_a \sim 10^{10}\,\mathrm{GeV}$ and a mass $m_a = \mathcal{O}(100)\,\mathrm{keV}$, both thermal and non-thermal production mechanisms contribute to the relic abundance, bearing in mind that the heavy neutrino mass $M_N$ can be treated as a free parameter~\cite{Greljo:2025suh}. At comparable temperatures, the decays of right-handed neutrinos generate the lepton asymmetry responsible for the BAU. Achieving the observed DM relic abundance, therefore, involves a non-trivial interplay between these mechanisms: one must simultaneously avoid overclosing the Universe while reproducing the measured DM density.

 %-----------------------------------------------------------------------------
\subsection{Leptogenesis} 
\label{sec:leptogenesis}

It is well-known that out-of-equilibrium decays of very heavy RHN can account for the baryon asymmetry of the Universe, via the so-called (thermal) leptogenesis mechanism~\cite{Fukugita:1986hr,Covi:1996wh,Covi:1996fm,Pilaftsis:1997jf,Buchmuller:1997yu}. Since, in our model, thermal leptogenesis takes place in the standard fashion, we will not go into the details on how the BAU is computed from RHN decays, and instead refer the reader to the vast dedicated literature in Refs.~\cite{Giudice:2003jh,Buchmuller:2004nz,Hagedorn:2017wjy}. However, we will outline several key features that are relevant for our scrutiny of the majoron parameter space.

We will consider thermal leptogenesis in two distinct regimes: the one-flavour approximation and the fully-flavoured regime. The former is typically valid when RHN decays occur at $T\gg 10^{12}\,{\rm GeV}$, where interactions mediated by the charged lepton Yukawas are out of equilibrium, and therefore the total lepton asymmetry is stored in a coherent flavour superposition. However, at lower temperatures, the tau and even the muon Yukawa interaction becomes efficient, effectively decohering the lepton asymmetries stored in their associated flavours, and the evolution of the total lepton asymmetry can no longer be tracked with a single flavour~\cite{Abada:2006ea}.

For the purposes of our analysis, the crucial role played by leptogenesis is to constrain the RHN mass scale. In the context of one-flavour thermal leptogenesis, Davidson and Ibarra~\cite{Davidson:2002qv} derived a lower bound on the lightest RHN mass, $M_{N_1}$, that needs to be satisfied in order to correctly reproduce the BAU. Depending on particular assumptions, such as the initial RHN population, this bound reads $M_{N_1}\gtrsim 10^{8}-10^{9}\,{\rm GeV}$. However, it has been shown that, upon including flavour effects, leptogenesis can be successful even for RHN significantly below the Davidson-Ibarra bound~\cite{Moffat:2018wke}. Since leptogenesis generally provides a lower bound on the RHN mass scale, in the context of majoron DM, it provides extremely complementary information to the parameter space. Indeed, both majoron decays into charged fermions/photons and DM production via freeze-in depend on a combination of $M_N$ and $f_a$. By effectively providing an independent handle on $M_N$, leptogenesis allows to chart the region in the $m_a-f_a$ parameter space where the model can account for both the BAU and the DM abundance while satisfying observational constraints.

Even though leptogenesis, majoron freeze-in, and decays are all sensitive to the mass scale of RHN, there is no exact correspondence between the three processes. For instance, in the presence of a hierarchical RHN spectrum, leptogenesis is mostly sensitive to the mass of the lightest RHN, $M_{N_1}$, whereas freeze-in production, as it becomes evident from Eq.~\eqref{eq:omega_fi_estimation}, will be controlled by the mass of the heaviest RHN that manages to thermalise. Alternatively, majoron decays to charged fermions depend on the whole RHN mass matrix (see App.~\ref{app:couplings}) and therefore will generally be sensitive to the heaviest RHN mass. As a consequence, for a given $M_{N_1}$, freeze-in production and decays could be strongly enhanced in the presence of strong hierarchies in the RHN spectrum, thus shrinking the available parameter space.

In order to avoid such scenarios and remain as conservative as possible, we always consider a mildly hierarchical RHN spectrum $M_{N_3}/M_{N_2}=M_{N_2}/M_{N_1}=2$, such that the only free mass scale is $M_{N_1}$. In this context, we explore the parameter compatible with thermal leptogenesis in the following fashion: in the one-flavour approximation, we follow Ref.~\cite{Giudice:2003jh}, whereas in the fully flavoured regime, we make uses of the code \texttt{ULYSSES}~\cite{Granelli:2020pim,Granelli:2023vcm} to compute the BAU and make use of its \texttt{MultiNest}~\cite{Feroz:2007kg,Feroz:2008xx,Feroz:2013hea} interface to explore the viable parameter space.

%-----------------------------------------------------------------------------
\section{Results}
\label{sec:analysis}

In order to characterize the window of the $m_a-f_a$ space in which majoron DM production is compatible with successful thermal leptogenesis and observational constraints, we carry out a dedicated exploration of the majoron parameter space. 
As shown in App.~\ref{app:couplings}, the seesaw parameters determine the structure of majoron couplings to photons and charged leptons, and the scale of RHN masses controls both the size of the aforementioned couplings, as well as leptogenesis and DM freeze-in production.

For our numerical scan, we employ the popular Casas-Ibarra parametrisation~\cite{Casas:2001sr}, which allows to scan the seesaw parameter space while correctly reproducing neutrino oscillation data. Within this choice, the full seesaw Lagrangian can be characterized with 11 free parameters\footnote{For simplicity and due to the precision with which they are measured, we fix the light neutrino mixing angles $\theta_{ij}$ and the mass splittings $\Delta m^2_{21}$ and $\Delta m_{31}^2$ to their best-fit values~\cite{Esteban:2024eli}.} $\boldsymbol{P}\equiv\left\{m_1,\delta,\alpha_{21},\alpha_{31},x_{1,2,3},y_{1,2,3},M_{N_1}\right\}$ (see App.~\ref{app:casas} for details). We recall that, since we are fixing the RHN mass hierarchy, only the lightest RHN mass is free. 

We define the viable window of $m_a-f_a$ parameter space in the following way (more details in App.~\ref{app:details-analysis}): 
\begin{enumerate}
    \item As computing the BAU from thermal leptogenesis is the most computationally expensive task in our analysis, our starting point is an ensemble of points in $\boldsymbol{P}$-space which is compatible with generating the correct BAU (see App.~\ref{app:lepto}).
    \item From each point $\boldsymbol{P}$, it is possible to predict the majoron couplings to photons and charged fermions normalised to its decay constant, $g_{aX}f_a$ with $X=f_{\rm SM},\gamma$.
    \item Using the irreducible contribution from misalignment, since each $\boldsymbol{P}$-point fixes the RHN masses and $\tilde{m}_i$, it is possible to find a trajectory in $m_a-f_a$ space in which the sum of the two contributions to the DM abundance (Eqs.~\eqref{eq:omega_misalignment} and~\eqref{eq:omega_fi_estimation}) yield the observed value $\Omega_{\rm DM}h^2\simeq 0.12$.\footnote{{Note that this condition immediately excludes any point in parameter space in which the majoron thermalises, since for masses $\gtrsim 100 \,\mathrm{eV}$ (see Refs.~\cite{Lattanzi:2007ux,Reig:2019sok}) its thermal abundance would overshoot the observed dark matter relic abundance.}} 
    \item  Over this trajectory, it is now possible to consistently apply DM decay constraints of Sec.~\ref{sec:constraints}, as their validity depends on DM being entirely made up of majorons.
    \item Since observational constraints place upper bounds on $g_{aX}$ as a function of $m_a$, these can be automatically converted to lower bounds on $f_a$, which are only applicable in the trajectory of the $m_a-f_a$ plane where the correct DM abundance can be reproduced. We remove all points of this trajectory in which the condition $f_a<f_a^{\rm DM-decay}(m_a)$ is satisfied. 
    \item 
    %Additionally, when $m_a\lesssim 5\,{\rm keV}$, we remove the points in which the fraction of warm DM $f_{\rm wDM}$ i.e., the contribution arising from freeze-in, exceeds the mixed wDM bounds (see \cref{sec:constraints}).
    {Additionally, for $m_a\lesssim5\,\mathrm{keV}$, we remove only those points for which the warm fraction $f_{\rm WDM}=\Omega^{\rm fi}_{\rm DM}/\Omega_{\rm DM}$, exceeds the mixed-wDM limit at the corresponding majoron mass~\cite{Tan:2024cek} (see \cref{sec:constraints}).}
\end{enumerate}

By repeating the above procedure for all $\boldsymbol{P}$-points and marginalising over them, it is possible to consistently map the \textit{largest} possible region of the $m_a-f_a$ in which leptogenesis 
and the correct DM abundance can be simultaneously achieved while respecting the bounds on DM decays and wDM. Our results are shown as the shaded green regions in Fig.~\ref{fig:parameter_space_post} and Fig.~\ref{fig:parameter_space_pre}, for the post-inflationary and pre-inflationary scenarios, respectively.

%****************************************
\subsection{Post-inflationary}

As discussed in Sec.~\ref{sec:darkmatter}, in the post-inflationary scenario, the initial misalignment angle is fixed to the average value $\theta_i\simeq \pi/\sqrt{3}$ and, given that $\Omega_{\rm DM}^{\rm mis}h^2\propto f_a^2m_a^{1/2}$, the allowed region for consistent majoron DM is bounded from above by misalignment overproduction, as shown by the dotted red line in Fig.~\ref{fig:parameter_space_post}. However, there is much theoretical evidence for the fact that the contribution coming from cosmic string radiation, albeit showcasing a similar parametric dependence, cf.~\cite{Reig:2019sok}, provides the larger contribution. In order to illustrate this, we show as a shaded red area the typical region of parameter space in which the whole DM can be produced via string radiation, using the estimates given in Ref.~\cite{Chathirathas:2025aan} for a temperature-independent ALP mass, which predict smaller $f_a$ by an $O(1-10)$ factor.

Alternatively, since the freeze-in contribution to the DM abundance~\eqref{eq:omega_fi_estimation} is inversely proportional to $f_a$ and leptogenesis requires a minimum RHN mass scale, $M_{N_1}$, the shaded green region is bounded from below by freeze-in overproduction, depicted by dotted blue lines in Fig.~\ref{fig:parameter_space_post}, depending on whether leptogenesis is computed in the one-flavour approximation or in the fully three-flavoured regime. Since flavour effects can help enhance the lepton asymmetry, three-flavour leptogenesis allows for lighter RHN and thus can accommodate more parameter space compatible with DM production. In fact, in the one-flavour regime, the upper bound from misalignment/strings and the lower bound from freeze-in and leptogenesis only allow for $m_a\lesssim \rm keV$, where wDM bounds constrain the abundance of frozen-in majorons to be a subdominant contribution to the total DM density. This can also be seen from the fact that below the wDM bound, the green region shrinks to coincide with the parameter space in which the total DM density can be produced non-thermally. {It is worth noting that a more hierarchical RHN mass spectrum than the one we have chosen would enhance the frozen-in abundance while not affecting the parameter space compatible with thermal leptogenesis. This would push the dotted blue lines in Figs.~\ref{fig:parameter_space_post} and~\ref{fig:parameter_space_pre} to higher $f_a$ values. Similarly, the loop-induced couplings to charged fermions would also benefit from an enhancement, extending the region excluded by DM decays to higher $f_a$. Alternatively, in the presence of nearly-degenerate RHN, the CP-asymmetry would be resonantly enhanced and therefore the scale of leptogenesis can be greatly reduced. However, we focus in this work only on the standard regime of thermal leptogenesis, since the degeneracies in the RHN spectrum necessary for resonant leptogenesis are difficult to motivate in our Type-I Seesaw without any underlying symmetry.}

Even though non-thermal and freeze-in DM production can, in principle, allow for majoron masses up to $m_a\sim 100\,\rm MeV$, the allowed region in Fig.~\ref{fig:parameter_space_post} reaches only up to $m_a\simeq 1\,\rm MeV$. This is a consequence of the interplay between leptogenesis, DM production and DM decay constraints: for majorons above the MeV scale, the decay channel $a\to e^+e^-$ opens, providing stringent constraints and since in a post-inflationary scenario, the scale $f_a$ needs to be relatively small to evade DM overproduction and, together with the fact that $M_{N_1}$ is bounded from below by leptogenesis, means that the $g_{ae}$ coupling cannot be arbitrarily small. In fact, we find that in none of the points of our numerical scan is $g_{ae}$ below the current observational constraints, effectively excluding majoron DM with $m_a>2m_e$. This implies that, unfortunately, neutrino telescopes (whose bounds are depicted in gray in Fig.~\ref{fig:parameter_space_post}) will not be able to probe our majoron setup, as they can only probe DM decays to neutrinos for $m_a\gtrsim \rm few\,MeV$. 

Lastly, we show in Fig.~\ref{fig:parameter_space_post} as dashed black lines the region that will be probed in the future, considering the forecasted sensitivities of the observatories outlined in Sec.~\ref{sec:constraints}. As just discussed, upcoming neutrino observatories, such as JUNO and HyperK, will not be able to access the relevant parameter space. On the other hand, apart from neutrinos, a sub-MeV majoron can only decay to a pair of photons. Consequently, the only observational facilities that will be able to probe the parameter space will be X-ray/gamma-ray observatories. In particular, as pointed out in sec.~\ref{sec:constraints}, we have included forecasts from a possible THESEUS mission~\cite{Thorpe-Morgan:2020rwc} and the Gamma-TPC telescope concept~\cite{Shutt:2025xvc}. The dashed black line in Fig.~\ref{fig:parameter_space_post} shows the region that these two proposals will be able to cover. In fact, we find that the improvement can only occur for $m_a\sim 0.1-1\,\rm MeV$, in which Gamma-TPC dominates the sensitivity. Unfortunately, THESEUS will have little to no impact due to it being sensitive to lighter majorons ($m_a\sim 1-100\,\rm keV$) and the majoron-photon coupling scaling as $g_{a\gamma}\propto m_a^2$ for masses below $m_e$.

{Since the green regions are obtained after marginalising over the full seesaw parameter space, their boundaries do not generally correspond to a unique seesaw configuration. Nevertheless, the freeze-in dominated boundary preferentially selects the lowest RHN mass scale compatible with successful leptogenesis, while the decay-dominated boundary favours smaller combinations of $M_D M_D^\dagger$ entering the loop-induced majoron couplings. The boundaries controlled by non-thermal production or by the $a\to\nu\nu$ lifetime are instead largely insensitive to the detailed RHN flavour structure.}

%****************************************
\subsection{Pre-inflationary}

As discussed in Sec.~\ref{sec:darkmatter}, the pre-inflationary scenario differs from its post-inflationary counterpart by the absence of DM production from topological defects, where the total non-thermal contribution arises from misalignment with the initial misalignment angle $\theta_i$ being a free parameter $|\theta_i|\in [0,\pi]$. In particular, since now it is possible to have $|\theta_i|\ll 1$, the target DM parameter space can be opened to higher $f_a$. In particular, as a reference, we consider values $|\theta_i|\in [10^{-3},1]$, and the region in which misalignment can be made compatible with the total DM abundance is shown as the shaded red region in Fig.~\ref{fig:parameter_space_pre}. 

Due to the higher $f_a$ values that are now allowed (provided that $|\theta_i|\ll1$) there is now a sizable window of parameter space in which 1F leptogenesis can be made compatible with DM production, which we show with a gray line.  Similarly, higher $f_a$ values also mean longer-lived majorons, opening parameter space for $m_a\geq 1\,\rm MeV$ for $f_a\gtrsim 10^{11}\,\rm{GeV}$, where bounds on the decay $a\to e^+e^-$ can be evaded in the three-flavoured regime of leptogenesis. This is not the case in the one-flavoured regime, which remains compatible only with sub-MeV majorons (unless $|\theta_i|\ll 10^{-3}$). This is due to the 
heavier RHN masses required in this scenario for leptogenesis, contrary to the fully-flavoured regime. 
%This is a consequence of the heavier RHNs which are required in this regime with respect to the fully-flavoured one. 
In fact, since majoron couplings to charged fermions scale with $M_N$ for a fixed $f_a$ (see App.~\ref{app:couplings}), larger values of $f_a$ are needed to evade the $a\to e^+e^-$ bound and open up parameter space for $m_a>\rm MeV$.

{Furthermore, as discussed in Sec.~\ref{sec:darkmatter}, this scenario unavoidably generates scale-free isocurvature perturbations, with the CMB providing a constraint on the energy scale of inflation~\eqref{eq:isocurvature_bound}, that translates into a bound on $T_{\rm rh}$. In particular, this leads to relatively small $T_{\rm rh}$ for the lowest $f_a$ values enclosed in our cosmological window, which can enter in conflict with our assumptions on the success of leptogenesis and DM freeze-in production. For instance, for $f_a=10^8\,\rm GeV$, we have $T_{\rm rh}\leq E_I\lesssim\sqrt{\theta_i}\,10^{11}\,{\rm GeV}$. However, we can still extend our window to lower $f_a$ by reducing the scale of the RHN taking part in leptogenesis and DM production. Indeed, the lower region of our cosmological window in Fig.~\ref{fig:parameter_space_pre} is populated by RHN below the Davidson-Ibarra bound, i.e., $M_{N_i}\ll10^{9}\,{\rm GeV}$, meaning that the above $T_{\rm rh}$ bound is always compatible with leptogenesis.}

Despite the allowed parameter space, up to $m_a\simeq 100\,\rm MeV$, the associated $f_a$ values are slightly above what current or future neutrino telescopes can probe, again underscoring how powerful the constraints on visible DM decays are, despite arising at the loop level for the majoron.

The future prospects of probing the allowed window are shown as black dashed lines in Fig.~\ref{fig:parameter_space_pre}. Since heavier masses are now allowed, the Gamma-TPC proposal can constrain more parameter space, as its sensitivity extends up to $m_a\sim 10\,\rm MeV$. As it can be observed from Fig.~\ref{fig:parameter_space_pre}, it will be able to exclude majorons above a few $\rm MeV$, except for a small window corresponding to $m_a\sim\mathcal{O}(10-100\,\rm MeV)$ that is, unfortunately, just beyond the sensitivity of next-generation neutrino observatories.

%-----------------------------------------------------------------------------

%-----------------------------------------------------------------------------
\section{Conclusions} 
\label{sec:conc}

The majoron offers one of the most minimal frameworks that can simultaneously and naturally explain the origin of neutrino masses, the abundance of dark matter~(DM), and the matter-antimatter asymmetry via thermal leptogenesis. In this work, we have investigated this connection in the singlet majoron model, focusing on the region in which DM is produced without thermalising the majoron sector, and the baryon asymmetry is generated through standard high-scale thermal leptogenesis.

The central point of our analysis is that these requirements are not independent. Successful thermal leptogenesis fixes, or at least strongly constrains, the right-handed neutrino mass scale. The same scale controls the irreducible freeze-in production of majorons and enters the loop-induced couplings responsible for visible DM decays. As a result, imposing leptogenesis removes otherwise flat directions in the majoron parameter space and turns the relic-density and decay constraints into predictive bounds in the $(m_a,\, f_a)$ plane. We have quantified this interplay through a scan of the seesaw parameter space compatible with neutrino data and successful leptogenesis, including both the one-flavour and fully flavoured regimes.

The main results, summarised in Figs.~\ref{fig:parameter_space_post} and~\ref{fig:parameter_space_pre} for the post- and pre-inflationary scenarios, cover the most prominent parameter space of this minimal model. In the post-inflationary case, the combination of misalignment production, the irreducible freeze-in abundance, warm-DM limits, and visible-decay constraints selects a relatively narrow window. In particular, the viable region is restricted to sub-MeV majoron masses, and scales above $f_a\gtrsim 10^8$~GeV, with the opening of the $a\to e^+e^-$ channel excluding heavier DM in our scan. The remaining uncertainty is dominated by the contribution from the string network associated with $\textrm{U}(1)_{B-L}$ breaking, whose normalization is still subject to the usual uncertainties in global-string simulations.

In the pre-inflationary case, the initial misalignment angle becomes a free parameter. This allows larger values of $f_a$ and correspondingly longer majoron lifetimes, opening viable parameter space up to masses of order $100\,\mathrm{MeV}$. Nevertheless, visible decay searches remain highly constraining, even though the relevant couplings arise only at loop level. In both cosmological histories, the viable cosmological window is therefore shaped by a non-trivial balance between cold non-thermal production, warm freeze-in production, successful leptogenesis, and indirect-detection limits.

Our results also clarify the experimental prospects. Neutrino telescopes probe majoron decays directly through the tree-level coupling to neutrinos, but the region compatible with thermal leptogenesis and DM stability largely lies outside their current and near-future reach. Instead, the most promising improvements arise from $X$- and gamma-ray searches for visible decays, especially in the MeV region, where projections of the Gamma-TPC~\cite{Shutt:2025xvc} concept can test part of the remaining parameter space. By contrast, missions targeting lighter keV-scale photons are less effective as the majoron-photon coupling is strongly suppressed when $m_a\lesssim 100$~keV.

Overall, the type-I majoron model remains a viable and predictive framework for linking neutrino masses, DM, and the baryon asymmetry. The surviving parameter space is limited but well motivated, and its further exploration will require both improved indirect searches for decaying DM and a sharper understanding of the cosmological history of the broken $\textrm{U}(1)_{B-L}$ symmetry, in particular the role of string radiation in the post-inflationary scenario. 

An intriguing complementary probe is the measurement of the stochastic gravitational-wave background produced along the DM~\cite{Obata:2026qwx} and asymmetry. Although gravitational-wave detectors cannot yet reach the signals expected from leptogenesis directly (see Appendix~\ref{app:GW} and references therein), the cosmic string network associated with the breaking of the $\mathrm{U}(1)_{B-L}$ symmetry could yield an observable background for appropriate values of the breaking scale (see e.g.~\cite{Chang:2019mza,DiBari:2023mwu}), offering a promising complementary window into the high-scale physics of this hermetic yet rich model.

%-----------------------------------------------------------------------------
\subsection*{Note added}
%-----------------------------------------------------------------------------
During the very last stage of writing of this manuscript, Ref.~\cite{Akita:2026gzk} appeared, addressing the same scenario with a complementary analysis. Our work provides an independent assessment of the viable parameter space, with a detailed scan of the seesaw parameters and a unified treatment of relic-density, leptogenesis, warm-DM, and decay constraints.

%-----------------------------------------------------------------------------
%-----------------------------------------------------------------------------
%\newpage
\subsection*{Acknowledgements}
The authors thank K.~Chathirathas, M.~Kaltschmidt, 
Y.~F.~Perez-Gonzalez, 
I.~Mart\'inez~Soler and 
J.~Turner 
for useful discussions and comments that improved the work.
 The work of DNT was supported by the Spanish MIU through the National Program FPU (grant number FPU20/05333) and by the Alexander von Humboldt Foundation.
 This article is based upon work from COST Action COSMIC WISPers CA21106,
supported by COST (European Cooperation in Science and Technology).

\newpage
\appendix
\onecolumngrid

%%%%%%%%%%%%%%%%%%%%%%%%%%%%%%%%%%%%%%%%%%%%%%%%
%%%%%%%%%%%%%%%%%%%%%%%%%%%%%%%%%%%%%%%%%%%%%%%%

%%%%%%%%%%%%%%%%%%%%%%%%%%%%%%%%%%%%%%%%%%%%%%%%%%%%%%%%%
%%%%%%%%%%%%%%%%%%%%%%%%%%%%%%%%%%%%%%%%%%%%%%%%%%%%%%%%%
\section{Majoron Couplings}
\label{app:couplings}
In this appendix, we collect the majoron couplings used to recast ALP constraints onto the majoron parameter space. The majoron couples at tree level only to neutral leptons, whereas its interactions with charged fermions and gauge bosons are generated radiatively. The leading couplings relevant for our phenomenological analysis, namely those to leptons, nucleons, and photons, are given by~\cite{Heeck:2019guh}
\begin{align}
    \mathcal{L}_{a-\textrm{SM}}=ia\left[\frac{1}{2}\bar{\nu} \frac{m^\nu}{f_a} \gamma_5 \nu +\bar{\ell}  \left(g^S_\ell +  g^P \gamma_5 \right)\ell +\bar n g_n \gamma_5 n+\bar p g_p \gamma_5 p -\frac{1}{4}g_{a\gamma\gamma}F\tilde{F}\right]\, ,
    \label{eq:couplings_init}
\end{align}
with
\begin{align}
    \label{eq:scalarLepton}
    g^S_{\alpha\beta}&=\frac{1}{16\pi^2 v^2 f_a}\left[(m_{\ell_\alpha}-m_{\ell_\beta}) (M_D M^\dagger_D)_{\alpha\beta} \right]\, ,\\
    \label{eq:pseudoscalarLepton}
    g^P_{\alpha\beta} &= \frac{1}{16\pi^2 v^2 f_a}\left[m_{\ell_\alpha}\delta_{\alpha\beta} \Tr{\left(M_D M^\dagger_D\right)} -(m_{\ell_\alpha}+m_{\ell_\beta}) (M_D M_D^\dagger)_{\alpha\beta}\right]\, ,\\
    \label{eq:nucleon_coupling}
    g_p &=\frac{-1.3 m_p}{16\pi^2 f_a}\frac{\Tr{\left(M_D M^\dagger_D\right)}}{v^2}\, , \quad
    % \label{eq:neutroncoupling}
    g_n =\frac{1.24 m_n}{16\pi^2 f_a}\frac{\Tr{\left(M_D M^\dagger_D\right)}}{v^2} \, , \\
     g_{a\gamma\gamma}&=\frac{\alpha_\textrm{em}}{8\pi^3 v^2 f_a}\left[
    \sum_l (M_D M_D^\dagger)_{ll}\,
    h\!\left(\frac{m_a^2}{4m_l^2}\right)+
    \operatorname{Tr}(M_D M_D^\dagger)
    \sum_f N_c^f Q_f^2 T_3^f\,
    h\!\left(\frac{m_a^2}{4m_f^2}\right)\right]\,,
    \label{eq:couplings_fin}
\end{align}
where $l$ runs over leptons, $f$ over all fermions, $N_c^{u,d} = 3$, $T^u_3 = \frac{1}{2} = -T^{d,l}_3$, $\{Q_l, Q_d, Q_u\} = \{-1, -\tfrac{1}{3}, +\tfrac{2}{3}\}$, and
\begin{equation}
    h(x) \equiv -\frac{1}{4x}\left(
    \log\!\left[1 - 2x + 2\sqrt{x(x-1)}\right]
  \right)^2 - 1=
  \begin{cases}
    \dfrac{x}{3} + \dfrac{8x^2}{45} + \dfrac{4x^3}{35}
      + \mathcal{O}(x^4), & x \to 0, \\[10pt]
    -1 + \dfrac{(\pi + i\log 4x)^2}{4x}
      + \mathcal{O}(x^{-2}), & x \to \infty.
  \end{cases}
\end{equation}
The proportionality of the majoron-fermions coupling to their masses is guaranteed by its ALP~(pNGB) nature, which becomes manifest in the derivative basis (see e.g. Ref.~\cite{Bonilla:2022qgm} for a pedagogical derivation).
For a majoron lighter than a MeV, $m_a\ll\text{MeV}$, the photon coupling simplifies to
\begin{equation}
    g_{a\gamma\gamma}\approx 
  -\frac{\alpha_\textrm{em}\, m_a^2}{12\pi}
  \sum_f N_c^f Q_f^2\,
  \frac{g_{aff}}{m_f^3}\,,
\end{equation}
where $f$ runs over all fermions, $Q_f$ is their electric charge, $N_c^f$ is the number of colours, $g_{aff}$ is the diagonal component of $g_{\alpha\beta}^S$ for each fermion. 
Notice that in this limit, the coupling to photons becomes proportional to $m_a^2$, and thus vanishes in the massless limit. 

Schematically, ignoring particle generation indices, for $m_a\ll\text{MeV}$, the scaling of the majoron's couplings goes as
\begin{align}
    &\mathcal{L}_{a-\textrm{SM}}\sim ia\left[g_{a\nu}m_\nu \bar{\nu}\nu+g_{a\ell}m_\ell\bar{\ell}  \ell -\frac{1}{4}g_{a\gamma}F\tilde{F}\right]\, ,\\
    &g_{a\nu}\sim \frac{1}{f_a}\,, \quad g_{a\ell}
    \sim \frac{1}{16\pi^2}\frac{m_\nu}{v^2}\frac{M_N}{f_a}\,,  \quad g_{a\gamma}
    \sim \frac{\alpha_\textrm{em}}{64\pi^3}\frac{m_\nu}{v^2}\frac{M_N}{f_a}\left(\frac{m_a^2}{m_e^2}\right)\,,
    \label{eq:estimates_couplings}
\end{align}
 where to recast in terms of the seesaw formula, we have to add the Majorana Yukawa $Y_N=M_N/f_a$. Analogous scaling holds for quarks, nucleons, and gluons.

Due to the seesaw relation between the parameters, there are several regimes in which different effective couplings dominate. For instance, taking $M_N/f_a=1$, the majoron coupling to neutrinos dominates over the electron coupling when $g_{a\nu}m_\nu \gtrsim g_{a\ell}m_e$, which gives
\begin{equation}
\label{eq:elec_vs_neutrino}
    f_a \lesssim \frac{16\pi^2 v^2}{m_e} \approx 2\times 10^{10}~\text{GeV},
\end{equation}
where, $v=246$~GeV. An analogous discussion holds for photons, and the coupling to neutrinos dominates when
The neutrino coupling dominates over the photon one when $g_{a\nu} m_\nu \gtrsim g_{a\gamma}$, giving
\begin{equation}
    f_a \lesssim \frac{64\pi^3}{\alpha_\textrm{em}}\frac{v^2 m_e^2}{m_a^2} \approx 4\times 10^{11}~\text{GeV} \times \left(\frac{100~\text{keV}}{m_a}\right)^2\,.
\end{equation}
Therefore, the parameter space shown in the main text lives in the region where a non-trivial interplay between the strength of couplings occurs.

%%%%%%%%%%%%%%%%%%%%%%%%%%%%%%%%%%%%%%%%%%%%%%%%%%%%%%%%%
%%%%%%%%%%%%%%%%%%%%%%%%%%%%%%%%%%%%%%%%%%%%%%%%%%%%%%%%%
\section{Casas-Ibarra Parametrisation}
\label{app:casas}

In order to efficiently scan the seesaw parameter space while correctly reproducing light neutrino masses and mixings, as measured from neutrino oscillation data~\cite{Esteban:2024eli}, we make use of the popular Casas-Ibarra parametrisation~\cite{Casas:2001sr}, in which the Dirac Yukawa can be expressed as:
\begin{equation}
    M_D=i\,U_{\rm PMNS}\sqrt{m_\nu^{\rm diag}}R^T \sqrt{M_N}
    \label{eq:ci}
\end{equation}
where $U_{\rm PMNS}$ is the usual active neutrino mixing matrix~\cite{Pontecorvo:1967fh,Maki:1962mu},
\begin{equation}
    U_{\rm PMNS}=
    \begin{pmatrix}
        1&0&0\\
        0&c_{23}&s_{23}\\
        0&-s_{23}&c_{23}
    \end{pmatrix}
    \begin{pmatrix}
        c_{13}&0&s_{13}e^{-i\delta}\\
        0&1&0\\
        -s_{13}e^{i\delta}&0&c_{13}
    \end{pmatrix}
    \begin{pmatrix}
        c_{12}&s_{12}&0\\
        -s_{12}&c_{12}&0\\
        0&0&1
    \end{pmatrix}
    \begin{pmatrix}
        1&0&0\\
        0&e^{i\tfrac{\alpha_{21}}2}&0\\
        0&0&e^{i\tfrac{\alpha_{31}}2}
    \end{pmatrix}\,,
\end{equation}
while $m_\nu^{\rm diag}={\rm diag}(m_1,m_2,m_3) $ and $M_N$ are diagonal matrices of mass eigenstates for light neutrinos and RHN, respectively. The $R$-matrix is a complex $3\times3$ orthogonal matrix (i.e. $RR^T=R^TR=1$), and can be parametrised with 3 complex angles $\omega_i=x_i+iy_i$:
\begin{equation}
    R=
    \begin{pmatrix}
        1&0&0\\
        0&c_{\omega_1}&s_{\omega_1}\\
        0&-s_{\omega_1}&c_{\omega_1}
    \end{pmatrix}
    \begin{pmatrix}
        c_{\omega_2}&0&s_{\omega_2}\\
        0&1&0\\
        -s_{\omega_2}&0&c_{\omega_2}
    \end{pmatrix}
    \begin{pmatrix}
        c_{\omega_3}&s_{\omega_3}&0\\
        -s_{\omega_3}&c_{\omega_3}&0\\
        0&0&1
    \end{pmatrix} \, ,
\end{equation}
where we have been using the notation $c_{\omega}=\cos\omega$, $s_\omega=\sin\omega$, $c_{ij}=\cos\theta_{ij}$ and $s_{ij}=\sin\theta_{ij}$. The total amount of free parameters that we will consider for all of our scans is the following: $U_{\rm PMNS}$ has 3 real mixing angles $\theta_{ij}$ (which we will fix to their best-fit values~\cite{Esteban:2024eli}) and three possible CP-violating phases (one Dirac phase, $\delta$, and possibly two extra Majorana phases $\alpha_{21}$ and $\alpha_{31}$). As for $m_{\nu}^{\rm diag}$, only the lightest neutrino mass eigenstate is free, as the other two are constrained by the measured mass splittings ($\Delta m_{21}^2$ and $\Delta m_{31}^2$). The $R$-matrix has $6$ real free parameters and the RHN masses constitute an additional $3$ parameters, but, as we have argued in the main text, we will fix the RHN mass hierarchy, such that only $M_{N_1}$ is a free parameter.

All in all, our seesaw sector has $11$ free parameters,
\begin{equation}
    \boldsymbol{P}\equiv\left\{m_1,\delta,\alpha_{21},\alpha_{31},x_{1,2,3},y_{1,2,3},M_{N_1}\right\}  \, ,
\end{equation}
over which we will scan in order to gauge which parameter regions are compatible with a successful thermal leptogenesis and to predict both majoron couplings to SM particles (see App.~\ref{app:couplings}) and the majoron abundances produced via freeze-in~\eqref{eq:omega_fi_estimation}.

Some key quantities for our analysis can be expressed as a function of this parametrisation. For example, the effective neutrino masses $\tilde{m}_i$, which determine the rate of RHN decays for leptogenesis, can be expressed as:
\begin{equation}
    \tilde{m}_i\equiv\frac{(M_D^\dagger M_D)_{ii}}{M_{N_i}}=\sum_{j=1}^3m_j |R_{ij}|^2 \, .
\end{equation}

Similarly, the matrix determining the structure of the majoron couplings to SM fermions and gauge bosons (see App.~\ref{app:couplings}), is given by:
\begin{equation}
    \frac{\left(M_DM_D^\dagger\right)}{vf}=U_{\rm PMNS} \sqrt{\frac{m_\nu^{\rm diag}}{v}}R^T\frac{M_N}{f_a}R^*\sqrt{\frac{m_\nu^{\rm diag}}{v}}U_{\rm PMNS}^\dagger  \, ,
\end{equation}
which explicitly showcases the parametric scaling of the loop-level majoron couplings with the RHN yukawas $Y_N\equiv \tfrac{\sqrt{2} M_N}{f_a}$.
%%%%%%%%%%%%%%%%%%%%%%%%%%%%%%%%%%%%%%%%%%%%%%%%%%%%%%%%%
%%%%%%%%%%%%%%%%%%%%%%%%%%%%%%%%%%%%%%%%%%%%%%%%%%%%%%%%%
\section{Majoron Thermalisation and Freeze-in}
\label{app:majoron_therm}

\begin{figure}[tbh]
\begin{tikzpicture}
\begin{feynman}
\def\xscale{0.25} %change to stretch in x
\def\yscale{0.3} %change to stretch in y
\vertex (i0) at (0*\xscale, 0*\yscale) {\( N\)};
\vertex (i1) at (0*\xscale, 5*\yscale) {\(N\)};
\vertex (i2) at (14*\xscale, 5*\yscale) {\(a\)};
\vertex (i3) at (14*\xscale, 0*\yscale) {\(a\)};
\vertex (i4) at (7*\xscale, 5*\yscale);
\vertex (i5) at (7*\xscale, 0*\yscale);
\diagram* {
(i1) -- [solid] (i4) -- [solid, edge label = \(N\)] (i5) -- [solid] (i0),
(i4) -- [scalar] (i2),
(i5) -- [scalar] (i3)
};
\end{feynman}
\end{tikzpicture}
\begin{tikzpicture}
\begin{feynman}
\def\xscale{0.25} %change to stretch in x
\def\yscale{0.3} %change to stretch in y
\vertex (i0) at (0*\xscale, 0*\yscale) {\(H\)};
\vertex (i1) at (0*\xscale, 5*\yscale) {\(L\)};
\vertex (i2) at (19*\xscale, 0*\yscale) {\(a\)};
\vertex (i3) at (19*\xscale, 5*\yscale) {\(N\)};
\vertex (i4) at (7*\xscale, 3*\yscale);
\vertex (i5) at (12*\xscale, 3*\yscale);
\diagram* {
(i1) -- [solid] (i4) -- [solid, edge label = \(N\)] (i5) -- [solid] (i3),
(i0) -- [scalar] (i4),
(i5) -- [scalar] (i2)
};
\end{feynman}
\end{tikzpicture}
\begin{tikzpicture}
\begin{feynman}
\def\xscale{0.25} %change to stretch in x
\def\yscale{0.3} %change to stretch in y
\vertex (i0) at (0*\xscale, 0*\yscale) {\(H\)};
\vertex (i1) at (0*\xscale, 6*\yscale) {\(N\)};
\vertex (i2) at (14*\xscale, 6*\yscale) {\(a\)};
\vertex (i3) at (14*\xscale, 0*\yscale) {\(L\)};
\vertex (i4) at (7*\xscale, 6*\yscale);
\vertex (i5) at (7*\xscale, 0*\yscale);
\diagram* {
(i1) -- [solid] (i4) -- [solid, edge label = \(N\)] (i5) -- [solid] (i3),
(i0) -- [scalar] (i5),
(i4) -- [scalar] (i2)
};
\end{feynman}
\end{tikzpicture}
\begin{tikzpicture}
\begin{feynman}
\def\xscale{0.25} %change to stretch in x
\def\yscale{0.3} %change to stretch in y
\vertex (i0) at (0*\xscale, 0*\yscale) {\(L\)};
\vertex (i1) at (0*\xscale, 6*\yscale) {\(N\)};
\vertex (i2) at (14*\xscale, 6*\yscale) {\(a\)};
\vertex (i3) at (14*\xscale, 0*\yscale) {\(H\)};
\vertex (i4) at (7*\xscale, 6*\yscale);
\vertex (i5) at (7*\xscale, 0*\yscale);
\diagram* {
(i1) -- [solid] (i4) -- [solid, edge label = \(N\)] (i5) -- [scalar] (i3),
(i0) -- [solid] (i5),
(i4) -- [scalar] (i2)
};
\end{feynman}
\end{tikzpicture}
    \caption{Representative Feynman diagrams leading to the possible thermalisation of the majoron and freeze-in relic abundance. The first diagram contributes only via the RH-Majorana Yukawa, while the other three also depend on the Yukawa of the seesaw. }
    \label{fig:FeynmanDiagrams}
\end{figure}

In this appendix, we analyse the possible thermalisation channels of the majoron with the SM thermal bath. If the majoron sector thermalises, majoron-mediated scatterings can keep the right-handed neutrinos in equilibrium for longer and thereby modify the standard high-scale thermal-leptogenesis dynamics~\cite{Vilja:1993uw,Gu:2009hn,Gu:2010ys,AristizabalSierra:2014uzi}. A thermalised majoron decouples while relativistic and therefore behaves as a hot relic; it cannot constitute the dominant DM component in the mass range of interest, since it would excessively suppress the formation of small-scale cosmological structure. Only very light majorons are compatible with this cosmological history, where non-thermal DM mechanisms set the relic abundance. In this work, we have studied the region that opens up if the majoron is never in thermal equilibrium with the SM. 

Direct couplings between the majoron and SM fields arise only after integrating out the RH neutrinos. For instance, the coupling to light neutrinos can be understood as descending from the higher-dimensional operator $
a\,(\overline{L_L^c}\,\tilde H^\ast)
(\tilde H^\dagger L_L)$. Similarly, the couplings to charged fermions, which are generated at one loop, also arise from operators involving Higgs insertions. As a consequence, the effective direct couplings of the majoron to SM fermions are only generated after electroweak symmetry breaking.

This observation is particularly relevant for the thermal history of the majoron. The usual SM thermalisation channels, such as two-to-two scatterings, inverse decays, and Primakoff processes, become active only once the Higgs acquires a non-zero vev. For axions and generic ALPs, these processes have been widely studied in the literature, see e.g.~\cite{Badziak:2024qjg,DEramo:2024jhn}. Making use of the results of Ref.~\cite{DEramo:2024jhn}, after rescaling the fermion and gauge-boson couplings by the naive estimates in \cref{eq:estimates_couplings}, one finds that the corresponding majoron interaction rates are far too suppressed to bring the majoron into thermal equilibrium with the SM bath. Inverse decays are also negligible in the high-scale majoron regime, for instance, for $a\to \nu \nu$ processes see Ref.~\cite{Escudero:2019gvw}.

Having argued that, in high-scale majoron models, the angular mode cannot thermalise through its direct SM interactions, we now turn to the seesaw sector. In the remainder of this appendix, we derive the conditions under which interactions with the RH neutrinos can bring the majoron into thermal equilibrium. We then compute the corresponding irreducible freeze-in contributions to the majoron abundance in the regime where thermalisation is not reached. 

The contribution from the processes $L_\alpha H\to a N_i$, $H N_i\to a L_\alpha$, $L_\alpha N_i\to a H$, and $N_iN_i \to aa$ {(drawn in \cref{fig:FeynmanDiagrams})} can be written in terms of the reaction densities
\begin{equation}
\gamma^{LH,HN,LN}_{i}(T)
=
\frac{T}{64\pi^4}
\int_{M_i^2}^{\infty} ds\,
\sqrt{s}\,
\widehat{\sigma}^{LH,HN,LN}_{i}(s)\,
K_1\!\left(\frac{\sqrt{s}}{T}\right) ,
\end{equation}
and
\begin{equation}
\gamma^{NN}_{i}(T)
=
\frac{T}{64\pi^4}
\int_{4M_i^2}^{\infty} ds\,
\sqrt{s}\,
\widehat{\sigma}^{NN}_{i}(s)\,
K_1\!\left(\frac{\sqrt{s}}{T}\right) .
\end{equation}
For the $LH$ channel, where the initial states are massless, the reduced cross section is
\begin{equation}
\widehat{\sigma}^{LH}_{i}(s)=2s\,\sigma^{LH}_{i}(s).
\end{equation}
Summing over lepton flavours, the cross section is
\begin{align}
\sigma^{LH}_{i}(s)
=
\sum_\alpha \sigma_{\alpha i}\!\left(L_\alpha H \to a N_i\right)
=&
\,2\,\frac{(Y_D^\dagger Y_D )_{ii}}{32\pi}
\frac{M_i^2}{f_a^2}
\frac{(s-M_{N_i}^2)^3}
{s^2\left[(s-M_{N_i}^2)^2+M_{N_i}^2\Gamma_i^2\right]}
\simeq\,2\,\frac{(Y_D^\dagger Y_D )_{ii}}{32\pi f_a^2}\frac{x_i-1}{x_i^2}\,,
\label{eq:LH_xsec}
\end{align}
where $x_i\equiv s/M_{N_i}^2$.
Here, the factor of 2 accounts for the electroweak degrees of freedom. The width entering the propagator can be estimated from the standard RH-neutrino decay rate,
\begin{equation}
\Gamma_i\simeq \Gamma(N_i \to  L H)
=
\frac{M_i}{8 \pi}(Y_D^\dagger Y_D)_{ii}
\simeq
\frac{M_i}{8 \pi}\frac{m_\nu M_i}{v^2} ,
\end{equation}
where in the last step we used the seesaw scaling $(Y_D^\dagger Y_D)_{ii}\sim m_\nu M_i/v^2$. Notice how in the last equality of  Eq.~\ref{eq:LH_xsec} we have dropped the decay width, since as long as $\Gamma_i/M_{N_i}\ll 1$ (which is always the case for our parameter space), its presence is numerically irrelevant.

The same coupling combination controls the crossed processes $H N_i\to aL_\alpha$ and $L_\alpha N_i\to aH$. 
 
Summing over lepton flavours, and using the same electroweak multiplicity convention as above, one obtains
\begin{align}
    \sigma_i^{ LN}(s)&=\sum_\alpha \sigma_{\alpha i}(L_\alpha N_i\to Ha)=2\frac{(Y_D^\dagger Y_D)_{ii}}{16\pi f_a^2}\frac{1+x_i(\log x_i-1)}{(x_i-1)^2}\,,\\
    \sigma_i^{ HN}(s)&=\sum_\alpha \sigma_{\alpha i}(H N_i\to L_\alpha a)=2\frac{(Y_D^\dagger Y_D)_{ii}}{16\pi f_a^2}\frac{x_i\log x_i}{(x_i-1)^2}\,,\,\,\,\,\,\, {\rm with}\,\,\,\,\,\, x_i\equiv \frac{s}{M_{N_i}^2}\,.
\end{align}
with the corresponding reduced cross sections given by:
\begin{equation}
    \widehat{\sigma}_i^{LN/HN}(s)=2(s-M_{N_i}^2)\sigma_i^{LN/HN}(s)
\end{equation}

The $t/u$-channel process $N_iN_i\to aa$ is parametrically different, since it is controlled directly by the majoron coupling to RH neutrinos and does not depend on the Dirac Yukawa coupling. The corresponding cross section is
\begin{equation}
\sigma^{NN}_{i}(s)
\equiv
\sigma^{NN}_{ii\to aa}(s)
=
\frac{M_i^4}{32\pi f_a^4}
\frac{1}{s\,\beta_i^2(s)}
\left[
\frac{1}{2}\log{\left(\frac{1+\beta_i(s)}{1-\beta_i(s)}\right)}
-\beta_i(s)
\right]
\,,
\qquad
\beta_i(s)=\sqrt{1-\frac{4M_i^2}{s}} .
\end{equation}
The reduced cross section is therefore
\begin{equation}
\widehat\sigma^{NN}_{i}(s)
=
2s\,\beta_i^2(s)\,\sigma^{NN}_{i}(s)
=
\frac{M_i^4}{16\pi f_a^4}
\left[
\frac{1}{2}\log{\left(\frac{1+\beta_i(s)}{1-\beta_i(s)}\right)}
-\beta_i(s)
\right]
\, .
\end{equation}

In addition, there can be $s$-channel diagrams mediated by the radial mode $\rho$. We neglect these contributions in the regime considered here. Indeed, we assume that the majoron never thermalises, which requires avoiding efficient production of the radial sector. Since $m_\rho\sim f_a$, this amounts to working in a regime where the relevant temperatures are below the radial threshold, or where the radial mode is otherwise not populated. Under this assumption, the dominant thermal production channels are $L H\to N_i a$, $H N_i\to aL$, $L N_i\to aH$, and $N_iN_i\to aa$. All in all, to quantify whether these interactions are efficient, we compare their rate with the Hubble expansion. We define the majoron interaction rate as the production rate normalised to the equilibrium number density of majorons,
\begin{equation}
\Gamma_a^{(\mathcal P)}(T)
\equiv
N_a^{(\mathcal P)}
\frac{\gamma_i^{\mathcal P}(T)}{n_a^{\rm eq}(T)} ,
\end{equation}
where $\gamma_i^{\mathcal P}$ is the reaction density for the process $\mathcal P=LH,HN,LN,NN$, and $N_a^{(\mathcal P)}$ denotes the number of majorons produced in each reaction. For example, $N_a^{(NN)}=2$ for $N_iN_i\to aa$, while $N_a^{(LH)}=N_a^{(HN)}=N_a^{(LN)}=1$. Thermalisation is expected whenever $\Gamma_a(T)/H(T)\gtrsim 1$, for some range of temperatures.

\begin{figure}[tbh]
    \centering
    \includegraphics[width=\linewidth]{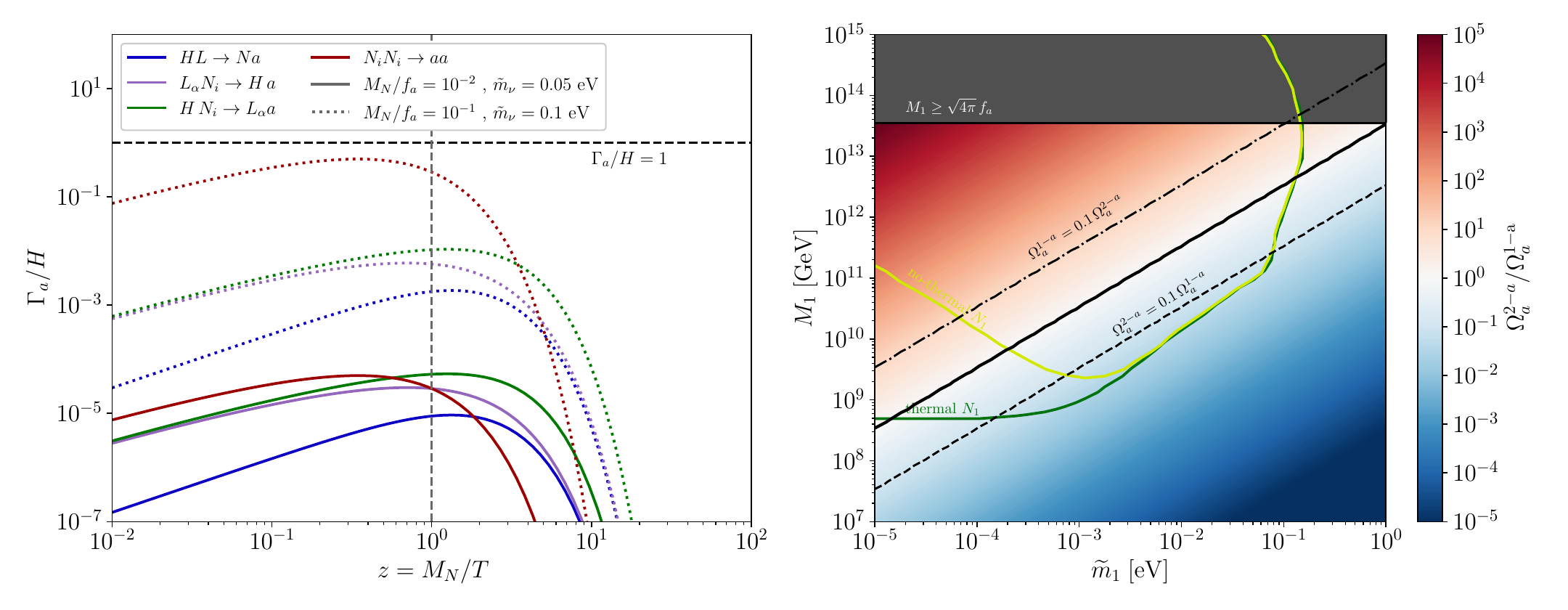}
    \caption{Thermalisation and freeze-in production of majorons from the seesaw sector. \textbf{Left:} interaction rates normalised to the Hubble expansion rate as a function of $z=M_N/T$, for the channels $HL\to Na$, $NN\to aa$, and the crossed processes $HN\to aL$, $LN\to aH$. The dotted and solid lines represent the two different benchmarks shown in the legend. \textbf{Right:} relative importance of the $NN\to aa$ and $HL\to Na$ production mechanisms in the $(\widetilde m_1,M_1)$ plane, for $f_a=10^{13}\,$GeV. The colour scale shows $\Omega_a^{2-a}/\Omega_a^{1-a}$, while the black contours indicate where each contribution accounts for $10\%$ of the observed DM abundance. The green and yellow contours delimit the regions where $N_1$ is thermally populated and where one-flavour thermal leptogenesis can be realised, respectively. The grey region corresponds to non-perturbative Majorana Yukawa coupling, $M_1\geq \sqrt{4\pi}\,f_a$.
}
    \label{fig:majoron_rates}
\end{figure}

In \cref{fig:majoron_rates} (left) we show the different interaction rates normalised to the Hubble rate, as a function of $z=M_N/T$. We observe that only the channel $N_iN_i\to aa$ can thermalise the majoron for sufficiently large Majorana Yukawa coupling, $Y_N\gtrsim 0.1$. The channels involving the Dirac Yukawa interaction are instead intrinsically more suppressed, as their size is tied to the light-neutrino masses through the seesaw relation. By contrast, production through $N_iN_i\to aa$ is controlled directly by the Majorana Yukawa coupling $Y_N$, and therefore does not inherit this additional seesaw suppression. However, the rate of this process scales as $Y_N^4$. As a result, even a moderate suppression of the Majorana Yukawa coupling rapidly reduces its efficiency, preventing the majoron from reaching thermal equilibrium.

Even if these processes cannot thermalise the majoron, they can generate an irreducible population of majorons via freeze-in. To compute the relic abundance, we can integrate the Boltzmann equation following Ref.~\cite{Blennow:2013jba}, finding:
\begin{align}
    \Omega_{\rm DM}^{2-a}h^2\simeq&\,
    0.12 \left(\frac{m_a}{100\,{\rm keV}}\right)
    \left(\frac{3.4\times 10^{10}\,{\rm GeV}}{f_a}\right)^4\sum_{i=1}^3\left(\frac{M_{N_i}}{1.6\times 10^{8}\,\rm GeV}\right)^3\,,\\
    \Omega_{\rm DM}^{1-a}h^2\simeq&\,0.12\left(\frac{m_a}{100\,{\rm keV}}\right)
    \left(\frac{3.4\times 10^{10}\,{\rm GeV}}{f_a}\right)^2\sum_{i=1}^3\left(\frac{\tilde{m}_i}{0.1\,\rm eV}\right)\left(\frac{M_{N_i}}{6.2\times 10^{8}\,\rm GeV}\right)^2\,,
\end{align}
where we have used the common definition in the leptogenesis literature $(Y_D^\dagger Y_D)_{ii}\equiv\frac{2}{v^2}\tilde{m}_iM_{N_i}$.

In \cref{fig:majoron_rates} (right), we show the regions of parameter space in which production through $N_iN_i\to aa$ and through $LH\to aN_i$ together with its crossed processes is most relevant. We also indicate the region where one-flavour thermal leptogenesis can successfully take place, following Ref.~\cite{Giudice:2003jh}. As expected from the parametric dependence of the rates, the $HL\to aN_i$ channel becomes comparatively more important when the Majorana Yukawa coupling is small and $\widetilde m_1$ is large, since in this regime the purely Majorana process $N_iN_i\to aa$ is suppressed while the Dirac-Yukawa-induced channels are enhanced.

\section{Treatment of Leptogenesis}
\label{app:lepto}

We sketch here the procedure for scanning the seesaw parameter space compatible with the correct BAU. As stated in the main text, we consider leptogenesis either in the one-flavour~(1F) approximation or in the fully-flavoured~(3F) regime.\\

\textbf{One-flavour approximation.} In this scenario, we closely follow Ref~\cite{Giudice:2003jh}, in which useful and easily-applicable results were presented after a careful evaluation of the Boltzmann equations governing leptogenesis. In particular, we make use of the fact that, when the RHN spectrum is hierarchical, the baryon-to-photon ratio can be expressed as~\cite{Giudice:2003jh}:
\begin{equation}
    \eta_B=-9.73\times 10^{-3}\epsilon_{N_1}\eta\,,
    \label{eq:bau_giudice}
\end{equation}
where $\epsilon_{N_1}$ is the CP-asymmetry in the decays of the lightest RHN, $N_1$, and $\eta$ is an efficiency factor. The former is given by
\begin{equation}
    \epsilon_{N_1}=\dfrac{1}{4\pi v^2}\sum_{j\neq1}\frac{{\rm Im}\left[\left(M_D^\dagger M_D\right)^2_{j1}\right]}{\left(M_D^\dagger M_D\right)_{11}}f\left(\frac{M_{N_j}^2}{M_{N_1}^2}\right)
\end{equation}
with $f(x)$ being a loop function defined in Eq.~(31) of Ref.~\cite{Giudice:2003jh}. Provided that $M_{N_1}\ll 10^{14}\,\rm GeV$ (as it is the case in our analysis), the efficiency only depends on the quantity $\tilde{m}_1$, defined as:
\begin{equation}
    \label{eq:parameters-lepto}\tilde{m}_1=\frac{\left(M_D^\dagger M_D\right)_{11}}{M_{N_1}}\,.
\end{equation}

The key observation is that, for a hierarchical RHN mass spectrum, the CP-asymmetry showcases the simple scaling $\epsilon_{N_1}\propto M_{N_1}$. Bearing this in mind, we perform our scan in the following way: we randomly scan in the space of $\boldsymbol{P}\equiv\left\{m_1,\delta,\alpha_{21},\alpha_{31},x_{1,2,3},y_{1,2,3},M_{N_1}\right\}$ keeping the RHN hierarchies fixed to $M_{N_3}/M_{N_2}=M_{N_2}/M_{N_1}=2$. From the Casas-Ibarra parametrisation (see App.~\ref{app:casas}), it is then possible to compute both $\tilde{m}_1$ (or, equivalently, $\eta$) and $\epsilon_{N_1}$. Since $\epsilon_{N_1}$ is the only quantity scaling with $M_{N_1}$, it is straightforward to re-scale $M_{N_1}$ in order to get the correct value for $\eta_{B}$ through Eq.~\eqref{eq:bau_giudice}, provided that $\epsilon_{N_1}$ has the correct sign. Using this strategy, it is easy to obtain a large amount of points compatible with the BAU without the use of dedicated sampling algorithms.\\

\textbf{Fully-flavoured regime}. In this scenario, the valuation of the BAU from the seesaw parameter space is a much more involved task. Indeed, this regime requires treating the lepton asymmetry as a matrix in flavour space, whose evolution captures the flavour-dependent wash-out and RHN decays, while also interpolating between the different flavoured regimes (which depend on the efficiency of the charged lepton Yukawa interactions)~\cite{Blanchet:2011xq,Moffat:2018wke}. For $M_{N_1}\ll 10^{9}\,\rm GeV$, we should expect both the $\tau$ and $\mu$ Yukawa interactions to be highly efficient, thus completely decohering the three flavours, in which case the density matrix equations become diagonal, effectively reducing to three Boltzmann equations for three flavoured CP asymmetries that evolve independently. However, as shown in Ref.~\cite{Moffat:2018wke}, even in this region of parameter space, there exist points in which the Boltzmann equations do not capture the full behaviour of the density matrix equations. Consequently, we always solve the density matrix equations for a single decaying RHN of mass $M_{N_1}$ using the publicly available \texttt{ULYSSES} code~\cite{Granelli:2020pim,Granelli:2023vcm} and its interface with \texttt{MultiNest}~\cite{Feroz:2007kg,Feroz:2008xx,Feroz:2013hea}, which allows for an easy (although expensive) exploration of the seesaw parameter space that is compatible with the correct value of the BAU. {We employ flat priors on the angles and phases entering the Casas-Ibarra parametrisation ($x_i,\,\delta,\,\alpha_{21},\,\alpha_{31}\in [0,2\pi]$, $y_i\in[-\pi,\pi]$) and a logarithmic prior on the lightest neutrino mass ($\log_{10}m_{1}/{\rm eV}\in[-4,-0.7]$) whereas the PMNS mixing angles are fixed to their best-fit values~\cite{Esteban:2024eli}. The parameter space in which enough lepton asymmetry can be produced shrinks with decreasing $M_{N_1}$, since $\epsilon_{N_1}\propto M_{N_1}$. For this reason, we do not scan in $M_{N_1}$, since we find that the sampling algorithm predominantly explores the upper part of the $M_{N_1}$ interval. Instead, we perform independent runs at fixed $M_{N_1}$ in the interval $\log_{10}M_{N_1}/{\rm GeV}\in [10^5,\,10^9]$ with steps of $\Delta \log_{10}M_{N_1}/{\rm GeV}=0.1$. Since \texttt{MultiNest} is called internally, we use the default configuration implemented in \texttt{ULYSSES}~\cite{Granelli:2020pim}. Given the high precision with which the baryon-to-photon ratio is measured, in the spirit of being conservative, we accept any point that lies within the $5\sigma$ interval as measured by Planck~\cite{Planck:2018vyg} (i.e., $10^{10}\eta=5.20-7.20$).}

In general, lowering the scale of thermal leptogenesis necessitates stronger RHN interactions with respect to the Hubble rate, entering the so-called \textit{strong wash-out} regime. In such a regime, one must be careful, as the rather large Yukawa couplings that are needed, together with the fact that the Casas-Ibarra parametrisation in Eq.~\eqref{eq:ci} is defined to automatically reproduce the light neutrino mass matrix, can easily lead to accepting parameter points in which the individual contributions of each RHN to $m_\nu$ are orders of magnitude above the scale of neutrino masses but conspire to cancel to give the correct result. In such cases, this fine-tuned cancellation is not perturbatively stable, and typically, including e.g. the 1-loop correction, spoils the fine-tuning and yields catastrophically large neutrino masses. Similarly, the overall size of the Yukawa couplings suggests that the tree-level implementation of the Casas-Ibarra parametrisation~\eqref{eq:ci} should be extended to include the one-loop contribution to $m_\nu$~\cite{Lopez-Pavon:2015cga}. This is achieved by making the following substitution in Eq.~\eqref{eq:ci}:
\begin{equation}
    M_N\longrightarrow \left(M_N^{-1} - \frac{M_N}{16\pi^2v^2}\left(\frac{\log\left(\tfrac{M_N^2}{m_H^2}\right)}{\left(\tfrac{M_N^2}{m_H^2}\right)-1}+3\frac{\log\left(\tfrac{M_N^2}{m_Z^2}\right)}{\left(\tfrac{M_N^2}{m_Z^2}\right)-1}\right)\right)^{-1}
\end{equation}
which is straightforward given that $M_N$ is diagonal.

We perform both the 1F and 3F analysis employing this modified parametrisation and, additionally, we asses how fine-tuned are the contributions to $m_\nu$ with a similar procedure as in Ref.~\cite{Moffat:2018wke}: for each $\boldsymbol{P}$ (see App.~\ref{app:casas}), we compute the 1-loop correction to the neutrino mass matrix, $m_\nu^{\rm 1-loop}$, perform a singular value decomposition~(SVD) and compare its singular values to those of the full $m_\nu$ (which simply correspond to the three neutrino masses $m_{1,2,3}$).
\begin{equation}
    \text{F.T.}\equiv\frac{\sum_{i=1}^3{\rm SVD}[m_\nu^{\rm 1-loop}]}{\sum_{i=1}^3 {\rm SVD}[m_\nu]}=\frac{\sum_{i=1}^3{\rm SVD}[m_\nu^{\rm 1-loop}]}{\sum_{i=1}^3 m_i}
\end{equation}
In the presence of large cancellations induced by the Casas-Ibarra parametrisation, the singular values of just the 1-loop part are much larger than $m_{1,2,3}$, yielding ${\rm F.T.\gg1}$. On the contrary, when there are no sizable cancellations within $m_\nu$, the loop contribution will be smaller than the total, simply by virtue of the $1/16\pi^2$ suppression. As such, after our initial scan, we will only keep $\boldsymbol{P}$-points which satisfy ${\rm F.T.}<0.2$. {This condition is specially relevant for our 3F leptogenesis scan. Indeed, after imposing it, our number of points decreases from $\sim 800\rm k$ to $\sim 300\rm k$. This is a consequence of our scan focusing on RHN masses below the Davidson-Ibarra bound, where the sampling algorithm tends to find solutions in which the CP-asymmetry is enhanced by unnaturally large Yukawa couplings, thus compensating for the relatively small $M_{N_1}$. On the contrary, in our 1F leptogenesis scan, which covers heavier RHN, $80\%$ of the initial points survive the fine-tuning condition.}

%%%%%%%%%%%%%%%%%%%%%%%%%%%%%%%%
\section{Details of the Analysis}
\label{app:details-analysis}
We provide here details on how we define the allowed green region in Figs.~\ref{fig:parameter_space_post} and~\ref{fig:parameter_space_pre}. As advertised in the text, the starting point is a scan of the seesaw parameter space that is compatible with leptogenesis, as presented in App.~\ref{app:lepto}. This scan consists on a set of points $\boldsymbol{P}\equiv\left\{m_1,\delta,\alpha_{21},\alpha_{31},x_{1,2,3},y_{1,2,3},M_{N_1}\right\}$ which we can use to predict the structure of majoron couplings and the freeze-in abundance. The goal is to employ this scan to chart the region of $m_a-f_a$ parameter space in which, if leptogenesis is successful, the majoron reproduces the DM abundance and survives observational bounds on decaying DM. Our analysis is carried in the following steps.\\

For each $\boldsymbol{P}$-point:

\begin{enumerate}    
    \item In order to determine the $m_a-f_a$ region in which the correct DM abundance can be generated, one needs to take into account both the non-thermal contribution (misalignment and string radiation, with the latter only being present in the post-inflationary scenario) and the contribution from freeze-in. The former only depends on $m_a$ and $f_a$ (once $\theta_i$ has been fixed), while the latter additionally depends on the RHN mass scale and on $\tilde{m}_i$, which are specified by the corresponding $\boldsymbol{P}$-point. Given this input, now it is possible to compute the total $\Omega_{\rm DM}h^2$ as a function of only $m_a$ and $f_a$. As such, one can solve the following equation:
    \begin{equation}
        \Omega_{\rm DM}^{\rm non-th}h^2+\Omega_{\rm DM}^{\rm fi}h^2=0.12 \, ,
    \end{equation}
    where the contributions from misalignment and freeze-in can be found in Eqs.~(\ref{eq:omega_misalignment},~\ref{eq:omega_fi_estimation}), respectively. The above equation can be rewritten in a more explicit fashion:
    \begin{equation}
        \left(\frac{f_a}{f_{\rm non-th}(m_a)}\right)^2+\left(\frac{f_{{\rm single}-a}(m_a)}{f_a}\right)^2+\left(\frac{f_{{\rm double}-a}(m_a)}{f_a}\right)^4=1 \, ,
        \label{eq:cardano_eq}
    \end{equation}
    where $f_{\rm non-th}$ is defined as the $f_a$ value, for a given $m_a$, in which the total DM abundance is generated via non-thermal production. In the case that non-thermal production is exclusively given by misalignment, the expression for $f_{\rm non-th}$ can be easily extracted from Eq.~\eqref{eq:omega_misalignment}:
    \begin{equation}
        f_{\rm non-th}(m_a)\simeq\frac{3.4\times 10^{10}\,\rm GeV}{\theta_i} \left(\frac{100\,\rm keV}{m_a}\right)^{1/4}\,,
    \end{equation}
    whereas $f_{{\rm single}-a}$ and $f_{{\rm double}-a}$ are defined similarly to $f_{\rm non-th}$, but for the single and double majoron freeze-in channels, respectively. They can be directly extracted from Eq.~\eqref{eq:omega_fi_estimation}
    \begin{equation}
    \begin{split}
        f_{{\rm single}-a}(m_a)&=3.4\times 10^{10}\,{\rm GeV} \left(\frac{m_a}{100\,\rm keV}\right)^{1/2}\left(\sum_{i=1}^3\left(\frac{M_{N_i}}{2.2\times 10^9\,\rm GeV}\right)^2\left(\frac{\tilde{m}_i}{0.1\,{\rm eV}}\right)\right)^{1/2}\\
        f_{{\rm double}-a}(m_a)&=3.4\times 10^{10}\,{\rm GeV}\left(\frac{m_a}{100\,\rm keV}\right)^{1/4}\left(\sum_{i=1}^3\left(\frac{M_{N_i}}{1.6\times 10^8\,\rm GeV}\right)^{3}\right)^{1/4} \, .
    \end{split}
    \end{equation}

    Eq.~\eqref{eq:cardano_eq} is a polynomial equation of third degree in $f_a^2$, which can be solved using Cardano's method~\cite{Cardano1545}. Whenever this polynomial has real solutions for $f_a^2$, it has two positive roots which correspond to two possible values of $f_a$ which, for a given $m_a$, can accommodate the correct DM abundance: the largest $f_a$ value corresponds to dominant non-thermal production, whereas the smallest root is a solution in which freeze-in dominates. Furthermore, there will be some $m_a$ value above which the corresponding $f_{\rm non-th}$, $f_{{\rm single}-a}$, and $f_{{\rm double}-a}$ render Eq.~\eqref{eq:cardano_eq} with no real solutions, in which case non-thermal production along with freeze-in overproduce DM $\forall f_a$. All in all, Eq.~\eqref{eq:cardano_eq} defines a trajectory in $m_a-f_a$ along which the majoron constitutes the total of the DM density.

    \item Over this trajectory, it is now consistent to apply observational bounds on decaying DM, as discussed in Sec.~\ref{sec:constraints}. In particular, the most relevant bounds we will consider are those stemming from visible majoron decays, mainly to photons, electrons, and muons. Even though we include decays to heavier particles, such as $\tau$ or $b$-quarks, these do not play any role in our target parameter space, as we find no viable majorons above the muon mass. These visible decays constrain the couplings presented in Eqs.~\ref{eq:couplings_init} to~\ref{eq:couplings_fin}. These can be predicted for each $\boldsymbol{P}$-point, up to an overall $1/f_a$ factor. It is therefore possible to recast the bounds on $g_{a\ell}$ and $g_{a\gamma\gamma}$ into a bound on $f_a$, which we will dub $f^{\rm vis-decay}$ by combining all the different visible channels. In practice, $f^{\rm vis-decay}$ is defined as:
    \begin{equation}
        f^{\rm vis-decay}(m_a)=f_0\sqrt{\sum_{\ell}\left(\frac{\left.g_{a\ell}\right\rvert_{f_a=f_0}}{g_{a\ell}^{\rm bound}(m_a)}\right)^2+\left(\frac{\left.g_{a\gamma\gamma}\right\rvert_{f_a=f_0}(m_a)}{g_{a\gamma\gamma}^{\rm bound}(m_a)}\right)^2}
    \end{equation}
    where $f_0$ is an arbitrary anchor value of $f_a$, which for example we choose $f_0=10^{10}\,\rm GeV$. The $m_a$-dependent bounds on the couplings are taken from the references outlined in Sec.~\ref{sec:constraints}.

    In addition to these bounds on visible DM decays, we include a bound on the lifetime $\tau_a>250\,\rm Gyr$ of the majoron, as obtained by a combination of CMB+LSS data~\cite{Simon:2022ftd}. This bound turns out to only be relevant in a region of our parameter space in which the total majoron decay width is dominated by the neutrino channel $a\to \nu\nu$. As such, the majoron lifetime can be easily computed from Eq.~\eqref{eq:neutrino_decay_width}, with the lightest neutrino mass taken from $\boldsymbol{P}$. This yields an additional lower bound on $f_a$, which we will dub $f^{\rm lifetime}$. Formally, it is defined as:
    \begin{equation}
        f^{\rm lifetime}(m_a)=2.5\times 10^{8}\,{\rm GeV}\left(\frac{m_a}{100\,\rm keV}\right)^{1/2}\left(\frac{\sum_{i=1}^3m_i^2}{2.5\times 10^{-3}\,{\rm eV^2}}\right)
    \end{equation}

    Lastly, the combined lower bound on $f_a$ obtained from DM decays is obtained by summing in quadrature $f^{\rm vis-decay}$ and $f^{\rm lifetime}$.
    \begin{equation}
        f^{\rm DM-decay}(m_a)=\sqrt{(f^{\rm vis-decay}(m_a))^2+(f^{\rm lifetime}(m_a))^2}
    \end{equation}

    \item We then remove all points in the $m_a-f_a$ trajectory reproducing the DM abundance that do not satisfy the condition {$f_a>f^{\rm DM-decay}$}.

    \item For the region corresponding to $m_a<5\,\rm keV$, we apply mixed wDM bounds, and remove points in the trajectory for which the frozen-in contribution is above the bounds mentioned in Sec.~\ref{sec:constraints}.
\end{enumerate}

We repeat this procedure for all $\boldsymbol{P}$-points and we marginalize over them, thus charting the biggest region of $m_a-f_a$ space in which both thermal leptogenesis is successful and the majoron can constitute the total DM abundance of the Universe, while surviving all astrophysical and cosmological constraints, which we dub the \textit{Majoron Cosmological Window}. The result is shown as the shaded green region in both Fig.~\ref{fig:parameter_space_post} and Fig.~\ref{fig:parameter_space_pre}.

%%%%%%%%%%%%%%%%%%%%%%%%%%%%%%%%%%%%%%
%%%%%%%%%%%%%%%%%%%%%%%%%%%%%%%%%%%%%%
\section{Gravitational Waves from Leptogenesis} 
\label{app:GW}
Besides conventional direct probes of majoron physics, one could hope to get complementary information directly from the neutrino sector. However, within Type-I seesaw scenarios, any hope for direct observation of neutrino physics is hopeless. Gravitational waves~(GWs) offer an exquisite complementary opportunity. The measurements of LIGO and VIRGO~\cite{LIGOScientific:2016aoc} have opened a new era for particle physics, with the possibility to directly access the early epochs dynamics of ultra-heavy particles, including RHNs. In fact, the Planck-suppressed couplings of GWs allow for the information to survive over a long time and be probed today. Several works have discussed the matter in the literature~\cite{Berbig:2023yyy,Borboruah:2025hai,Chianese:2025mll,Murayama:2025thw}. We report here some of the main findings, and most importantly, infer consequences and complementary avenues for Majoron physics.

\bigskip
\begin{figure}
    \centering
    \includegraphics[width=0.5\linewidth]{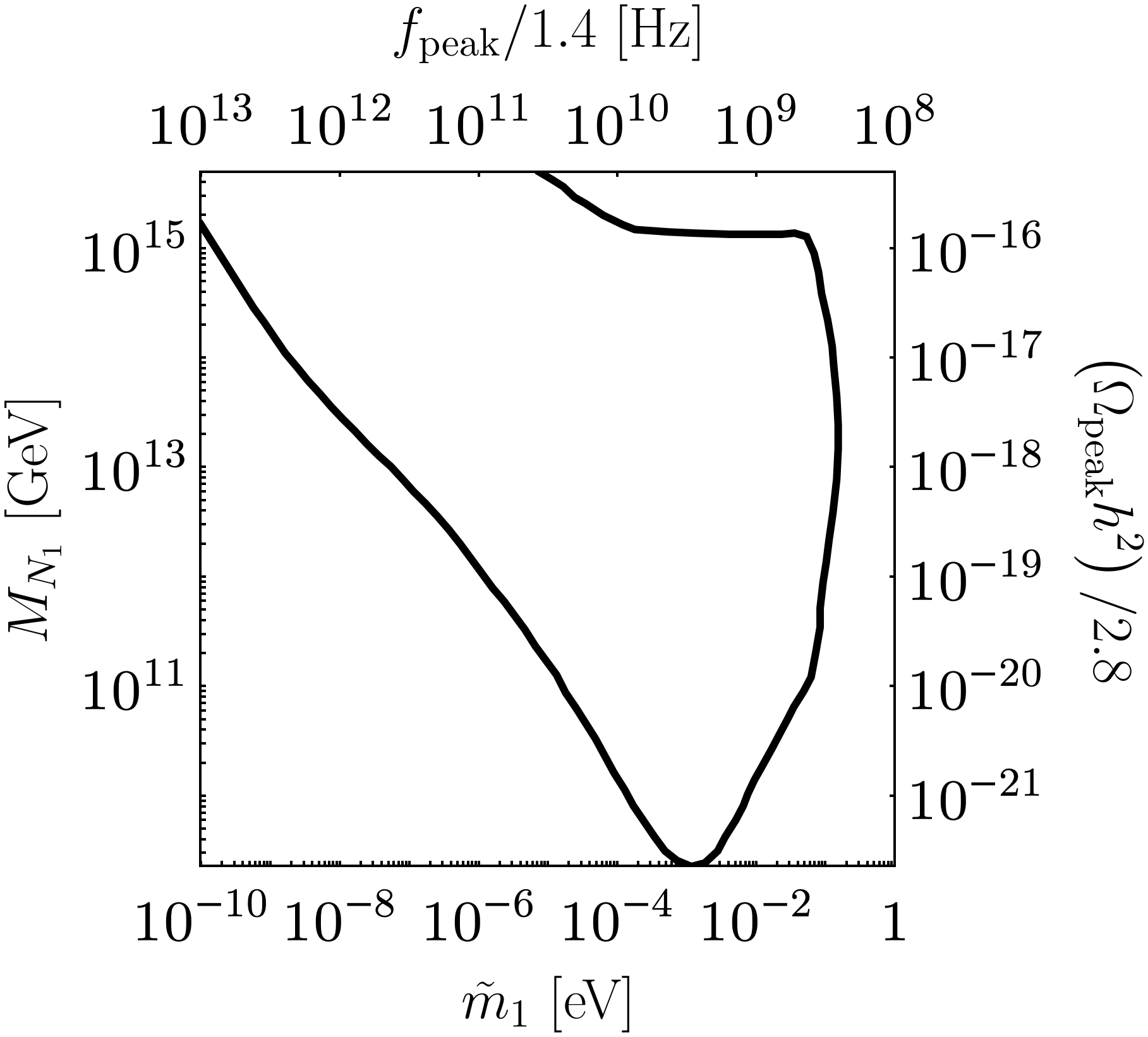}
    \caption{Parameter space available (within the black contour) for thermal leptogenesis in relation to the peak and abundance of the GWs spectrum.}
    \label{fig:leptowaves}
\end{figure}
A thermal bath of heavy RHNs, followed by their decay, can affect GW signals mainly in two ways:
\begin{enumerate}
    \item GWs can be emitted via bremsstrahlung directly during the decay of the RHN. The peak and magnitude of the signal are directly related to the mass and Yukawa of the lightest RHN.
    \item If the RHN does not decay instantaneously, they can induce an early matter-dominated epoch, thus affecting the SM prediction for the stochastic cosmic gravitational microwave background~(CGMB). The modification thus still bears information on the lightest RHN parameters.
\end{enumerate}
As noted in Ref.~\cite{Murayama:2025thw}, the simultaneous measurement of both signals would unambiguously point to this scenario and allow for measuring $Y\equiv \sqrt{(Y_DY_D^\dagger)_{11}}$ and $M_{N_1}$.
From the direct measurement of the peak position and magnitude, $(f_\text{gw}^\text{peak},h^2\Omega_\text{gw}(f_\text{gw}^\text{peak}))$ of the GWs spectrum from RHNs decay, one can estimate~\cite{Murayama:2025thw}
\begin{align}
    &M_{N_1}\approx 10^{15}~\text{GeV}\times\left[\frac{h^2\Omega_\text{gw}(f_\text{gw}^\text{peak})}{2.8\times 10^{-16}}\right]^{1/2}\,, \quad
     Y \approx 10^{-4}\times\left[\frac{h^2\Omega_\text{gw}(f_\text{gw}^\text{peak})}{2.8\times 10^{-16}}\right]^{1/4}\times\left[\frac{7.9\times 10^{12}~\text{Hz}}{f_\text{gw}^\text{peak}}\right]\,.
\end{align}
The GW details of the peak are therefore in a one-to-one correspondence with the parameters defined in Eq.~\eqref{eq:parameters-lepto} since
\begin{equation}
    \tilde{m}_1\approx 3\times 10^{-10}~\text{eV}\times\left[\frac{7.9\times 10^{12}~\text{Hz}}{f_\text{gw}^\text{peak}}\right]^2\,.
\end{equation}
The relation between the GWs parameters and the region of successful leptogenesis can be seen in Fig.~\ref{fig:leptowaves}. However, as argued in Ref.~\cite{Murayama:2025thw}, in the near future, it is therefore unlikely to be able to constrain leptogenesis. The measurements of high-frequency GWs ($f_\text{gw} \gtrsim 10~$~kHz) are extremely challenging (see Ref.~\cite{Aggarwal:2025noe} for a review), and new technological breakthroughs are needed to probe signals with such a small relic abundance. 

%-----------------------------------------------------------------------------
%-----------------------------------------------
%-----------------------------------------------------------------------------
%%
% Produces the bibliography via BibTeX.
\bibliographystyle{BiblioStyle}
\bibliography{refs}

%%%%%%%%%%%%%%%%%%%%%%%%%%%%%%%%%%%%%%%%%%%%%
%%%%%%%%%%%%%%%%%%%%%%%%%%%%%%%%%%%%%%%%%%%%%

%-----------------------------------------------------------------------------

\end{document}